\documentclass[aps,pra,twocolumn,superscriptaddress]{revtex4-2}
\usepackage{float}
\usepackage{graphicx}
\usepackage{dsfont}
\usepackage[latin1]{inputenc}
\usepackage{version} 
\usepackage{color}
\usepackage{amssymb}
\usepackage{amsmath, scalerel}
\usepackage{braket}
\usepackage{bm}
\usepackage{upgreek}
\usepackage{colortbl}
\usepackage{graphicx}
\usepackage{dcolumn}
\usepackage{bm}
\usepackage{bigints}
\usepackage{xcolor}  
\usepackage[pagebackref=false]{hyperref} 
\usepackage{enumitem}

\definecolor{jrp}{rgb}{1,0,0}

\definecolor{rim}{rgb}{0,1,0}

\definecolor{pr}{rgb}{0.7,0,0}

\definecolor{mjg}{rgb}{.08,.05,.8}

\definecolor{yyl}{rgb}{0.7,0.0,0.1}

\newcommand{\delete}[1]{{}} 
\colorlet{shadecolor}{gray!40}

\hypersetup{
    bookmarks=true,         
    unicode=false,          
    pdftoolbar=true,        
    pdfmenubar=true,        
    pdffitwindow=false,     
    pdfstartview={FitH},    
    pdftitle={dissipative_stablization},    
    pdfauthor={X. Mi},     
    pdfkeywords={keyword1} {key2} {key3}, 
    pdfnewwindow=true,      
    colorlinks=true,       
    linkcolor=red,          
    citecolor=blue,        
    filecolor=magenta,      
    urlcolor=cyan,           
}

\makeatletter

\makeindex

\begin{document}
\title{Stable Quantum-Correlated Many Body States through Engineered Dissipation}

\newcommand\Google{\affiliation{Google Research, Mountain View, CA, USA}}

\newcommand{\Geneva}{\affiliation{Department of Theoretical Physics, University of Geneva, Quai Ernest-Ansermet 30, 1205 Geneva, Switzerland}}

\newcommand{\UMass}{\affiliation{Department of Electrical and Computer Engineering, University of Massachusetts, Amherst, MA, USA}}

\newcommand{\AU}{\affiliation{Department of Electrical and Computer Engineering, Auburn University, Auburn, AL, USA}}

\newcommand{\CQC}{\affiliation{QSI, Faculty of Engineering \& Information Technology, University of Technology Sydney, NSW, Australia}}

\newcommand{\UCR}{\affiliation{Department of Electrical and Computer Engineering, University of California, Riverside, CA, USA}}

\newcommand{\CU}{\affiliation{Department of Chemistry, Columbia University, New York, NY, USA}}

\newcommand{\UoCA}{\affiliation{Department of Physics and Astronomy, University of California, Riverside, CA, USA}}

\author{X. Mi} \email[Corresponding author: ]{mixiao@google.com} \Google
\author{A. A. Michailidis}\Geneva
\author{S. Shabani}\Google
\author{K. C.~Miao}\Google
\author{P. V.~Klimov}\Google
\author{J. Lloyd}{\Geneva}
\author{E. Rosenberg}\Google
\author{R. Acharya}\Google
\author{I. Aleiner}\Google
\author{T. I. Andersen}\Google
% ...

\begin{abstract}
Engineered dissipative reservoirs have the potential to steer many-body quantum systems toward correlated steady states useful for quantum simulation of high-temperature superconductivity or quantum magnetism. Using up to 49 superconducting qubits, we prepared low-energy states of the transverse-field Ising model through coupling to dissipative auxiliary qubits. In one dimension, we observed long-range quantum correlations and a ground-state fidelity of 0.86 for 18 qubits at the critical point. In two dimensions, we found mutual information that extends beyond nearest neighbors. Lastly, by coupling the system to auxiliaries emulating reservoirs with different chemical potentials, we explored transport in the quantum Heisenberg model. Our results establish engineered dissipation as a scalable alternative to unitary evolution for preparing entangled many-body states on noisy quantum processors.
\end{abstract}

\maketitle

A major effort in quantum simulation and computation is devising scalable algorithms for preparing correlated states, such as the ground state of interacting Hamiltonians. On analog quantum simulators, states are often prepared via adiabatic unitary evolution from an initial Hamiltonian to a desired Hamiltonian \cite{BlochColdAtoms,BlattReview12,Altman_PRXQ_2021}. On digital quantum processors supporting more flexible unitary dynamics, variational quantum algorithms have also gained popularity in recent years \cite{Cerezo2021}. Both methods, however, have inherent limitations: Adiabatic state preparation is fundamentally difficult across quantum phase transitions where the many-body energy gaps close, whereas variational quantum algorithms involve large optimization overheads and are challenged by the so-called barren plateaus \cite{McClean2018}. The lifetimes of states prepared through unitary evolution are also limited by the coherence times of physical qubits, hindering their use as basis for noise-biased qubits \cite{Nori2014} or topological quantum computation \cite{KITAEV20032}.

\begin{figure}[t!]
    \centering
    \includegraphics[width=\columnwidth]{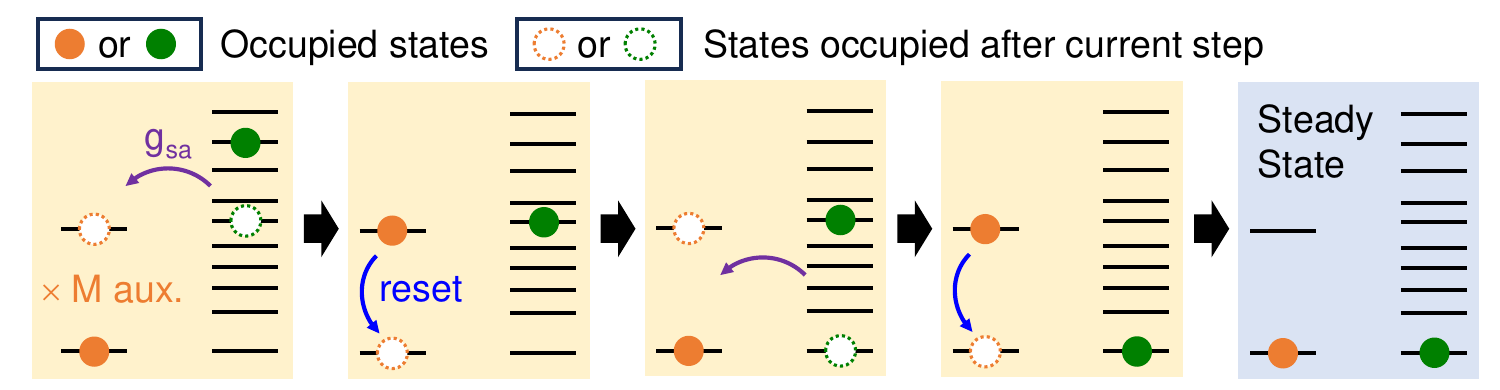} 
    \caption{Dissipative cooling of a many-body system (green dot) to its ground state via a reservoir comprising $M$ auxiliary two-level systems (orange dot), schematically illustrated as a sequence of steps.}
    \label{fig:1}
\end{figure}

An alternative and more robust route toward quantum state preparation is through engineered dissipation~\cite{Plenio_PRA_1999, Kraus2008, Diehl2008, Verstraete:2009vj, Hakan_PRX_2016, Harrington2022}. In such schemes, the quantum system is coupled to a dissipative reservoir that is repeatedly entangled with the system and projected to a chosen state. Over time, the system is steered toward a steady state of interest by the reservoir. A concrete example is dissipative cooling (Fig.~\ref{fig:1}). Here the reservoir is represented by $M$ two-level ``auxiliary'' qubits, each with an energy splitting close to the energy of low-lying excitations of a many-body quantum system \cite{Weimer_cooling_2020, Polla_PRA_2021}. The entangling operation, having a rate $g_\text{sa}$, transfers excitations from the system into the auxiliaries, which are then removed via controlled dissipation (``reset'') that brings the auxiliaries to their ground states. The process therefore cools the system toward its ground state. Even after the completion of the cooling process, the continued reset cycles of the
auxiliaries stabilize the cooled state against environmental decoherence, extending its lifetime beyond the coherence times of physical qubits.

\begin{figure*}[t!]
    \centering
    \includegraphics[width=1.333\columnwidth]{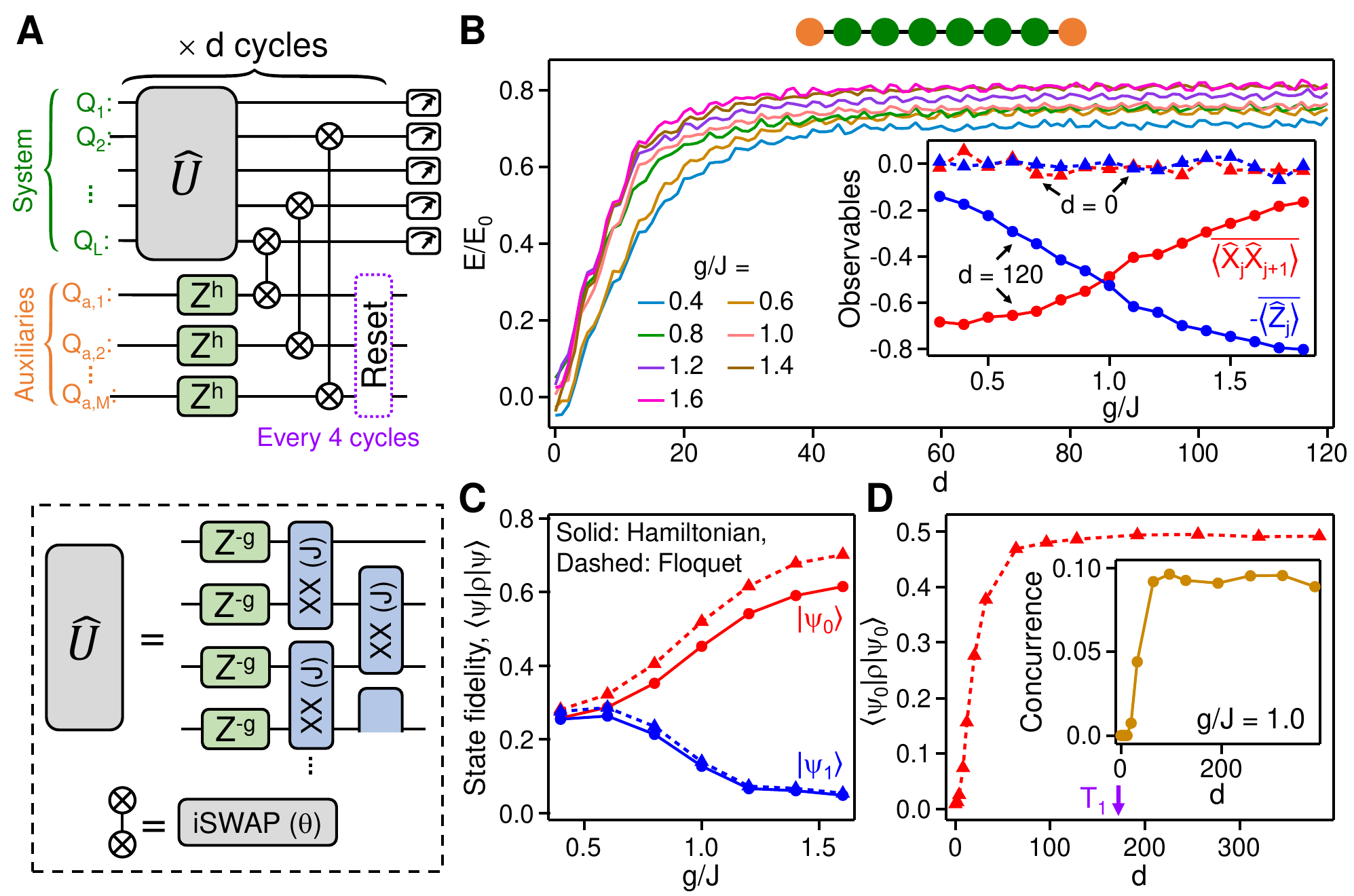} 
    \caption{{Dissipative cooling and stabilization of a 1D TFIM.} (A) A periodic quantum circuit used to implement dissipative cooling on a quantum processor. Here the $XX (J)$ and iSWAP($\theta$) gates are composed from tunable CPHASE and fermionic simulation (fSim) gates (see SM). (B) $E/E_0$ as a function of $d$ for different relative transverse field strengths, $g / J$. Here $E$ is the experimentally obtained energy and $E_0$ is the ground state energy. Inset shows the site-averaged observables $\overline{\braket{\hat{X}_j \hat{X}_{j + 1}}}$ and $-\overline{\braket{\hat{Z}_j}}$ as functions of $g / J$, measured at $d = 0$ and $d = 120$. (C) Fidelity of steady state ($d = 100$) with respect to the ground state of the TFIM, $\ket{\psi_0}$ (red), and the first excited state $\ket{\psi_1}$. Data are computed from experimental 6-qubit density matrices $\rho$. Here the solid lines correspond to eigenstates of the TFIM Hamiltonian and the dashed lines correspond to the Floquet eigenstates of the cycle unitary $\hat{U}$. (D) Floquet ground state fidelity $\bra{\psi_0} \rho \ket{\psi_0}$ as a function of $d$ for $g/J = 1.0$. Inset shows site-averaged nearest-neighbor concurrence as a function of $d$. The typical single-qubit $T_1 = 22$ $\mu$s corresponds to $d \approx 170$.}
    \label{fig:2}
\end{figure*}

Past experimental works have demonstrated the dissipative preparation of few-qubit states of trapped ions~\cite{Barreiro2011, Wineland2013} and superconducting qubits~\cite{Devoret:2013ve}, as well as an 8-qubit Mott insulator state in an analog quantum simulator~\cite{MaSchuster2019}. Dissipative preparation of many-body quantum states, however, has remained experimentally challenging due to increased environmental decoherence which threatens to overwhelm the impact of the auxiliaries. Open questions also remain on whether dissipatively prepared states with more than a few qubits in fact possess any non-classical characteristics. Dissipatively preparing many-body states and measuring their quantum correlation or entanglement entropy are therefore crucial for assessing the practical importance of engineered dissipation to current quantum hardware.

In this article, we report the preparation of many-body quantum states via dissipative cooling on a superconducting transmon quantum processor \cite{Acharya2023}. We provide experimental evidence of entanglement and long-range quantum correlations in the steady state, and demonstrate a favorable scaling of dissipative state preparation over system sizes when compared to unitary evolution algorithms. Furthermore, we extend the use of engineered dissipation beyond cooling and explore the non-equilibrium physics arising from coupling a many-body quantum system to two different reservoirs. This work is enabled by two technical advances: (i) Continuously tunable quantum gates with simultaneously operated two-qubit gate fidelities reaching 99.7\% in 1D and 99.6\% in 2D. Details of gate calibration are described in the Supplementary Materials (SM). (ii) A fast reset protocol comparable to unitary gates in duration, which reduces errors from qubit idling \cite{McEwen2021, Miao_Leakage_2022}.

We first perform a benchmarking experiment using a 6-site 1D transverse-field Ising model (TFIM) connected to two auxiliaries at the edges. The 1D TFIM is chosen since it is analytically solvable and has a quantum-entangled ground state. The Hamiltonian describing the system is:
\begin{equation}
\hat{H}_\text{TFIM} = - g\sum_{j=1}^{L} \hat{Z}_j + J\sum_{j=1}^{L - 1} \hat{X}_j \hat{X}_{j + 1}
\end{equation}
Here $\hat{X}$ and $\hat{Z}$ are Pauli operators whereas $g$ and $J$ denote control parameters. For $J > 0$, the model exhibits two quantum phases: an antiferromagnetic phase ($g/J < 1$) with two nearly degenerate ground states ($\sim \ket{+-} ^{\otimes L/2} \pm \ket{-+} ^{\otimes L/2}$, where $\ket{\pm} = \ket{0} \pm \ket{1} $) and a paramagnetic phase ($g/J > 1$) with a unique ground state ($\sim \ket{0} ^{\otimes L}$). At the critical point $g/J = 1$, the ground state is most entangled, having an entanglement entropy that grows logarithmically with subsystem size, and quantum correlations that decay as a power law over distance.

The dissipative cooling described in Fig.~\ref{fig:1} is implemented via $d$ cycles of a periodic quantum circuit on a system of qubits, $Q_1$ through $Q_{\rm L}$ (Fig.~\ref{fig:2}A). The quantum circuit includes a Trotter-Suzuki approximation of the time-evolution operator $\hat{U}_\text{TFIM} (\frac{\pi}{2}d) \approx \hat{U}^d$, where:
\begin{equation}
\hat{U} = e^{-\frac{i \pi J}{2} \sum_{j=1}^{L - 1} \hat{X}_j \hat{X}_{j + 1}} e^{\frac{i \pi g}{2} \sum_{j=1}^{L} \hat{Z}_j}.
\end{equation}
Unless otherwise stated, we use $J = 0.25$ for $g/J < 0.6$ and $J = 0.2$ for $g/J \geq 0.6$. Within every cycle, each auxiliary ($Q_{\rm a,1}$ through $Q_{\rm a,M}$) is also rotated with a phase gate $Z^h$, where the exponent $h$ effectively controls its energy splitting as illustrated in Fig.~\ref{fig:1}. Lastly, the auxiliaries are coupled to the system via a partial iSWAP gate with a tunable angle $\theta$, iSWAP($\theta$) = $e^{i\frac{\theta}{2} (\hat{X}\hat{X} + \hat{Y} \hat{Y})}$, where $\hat{Y}$ is another Pauli operator \cite{ziman_collision_2005}. Here $\theta$ controls the system-reservoir coupling $g_\text{sa}$ in Fig.~\ref{fig:1}. The auxiliaries are reset every 4 circuit cycles to allow sufficient time for energy exchange between the system and the reservoir \cite{Weimer_cooling_2020}. In the SM, we present additional experimental characterization that is used to determine the optimal values of $h$ and $\theta$. To demonstrate that our protocol may be applied to any initial state, all initial states used in the TFIM experiments are scrambled states prepared using random circuits having 50 cycles of CZ and single-qubit gates \cite{Boixo2018}.

We begin by characterizing the energy of the system, $E = \braket{\hat{H}_\text{TFIM}}$, as a function of $d$ and across the two phases of the TFIM. Figure~\ref{fig:2}B shows the time-dependent $E/E_0$ for different values of $g/J$, where $E_0$ is the ground state energy. We observe that $E/E_0$ increases from 0 at $d = 0$ to a stable value at $d > 50$, which ranges from 0.7 deep in the antiferromagnetic phase ($g/J = 0.4$) to 0.8 deep in the paramagnetic phase ($g/J = 1.8$).  The site-averaged observables, $\overline{\braket{\hat{X}_j \hat{X}_{j + 1}}}$ and $-\overline{\braket{\hat{Z}_j}}$, show an expected crossing around the critical point of the TFIM, $g/J = 1.0$ (inset to Fig.~\ref{fig:2}B). These data are preliminary indications that our dissipative cooling protocol is robust across quantum phase transitions, where the nature of the ground state qualitatively changes.

\begin{figure*}[t!]
    \centering
    \includegraphics[width=2\columnwidth]{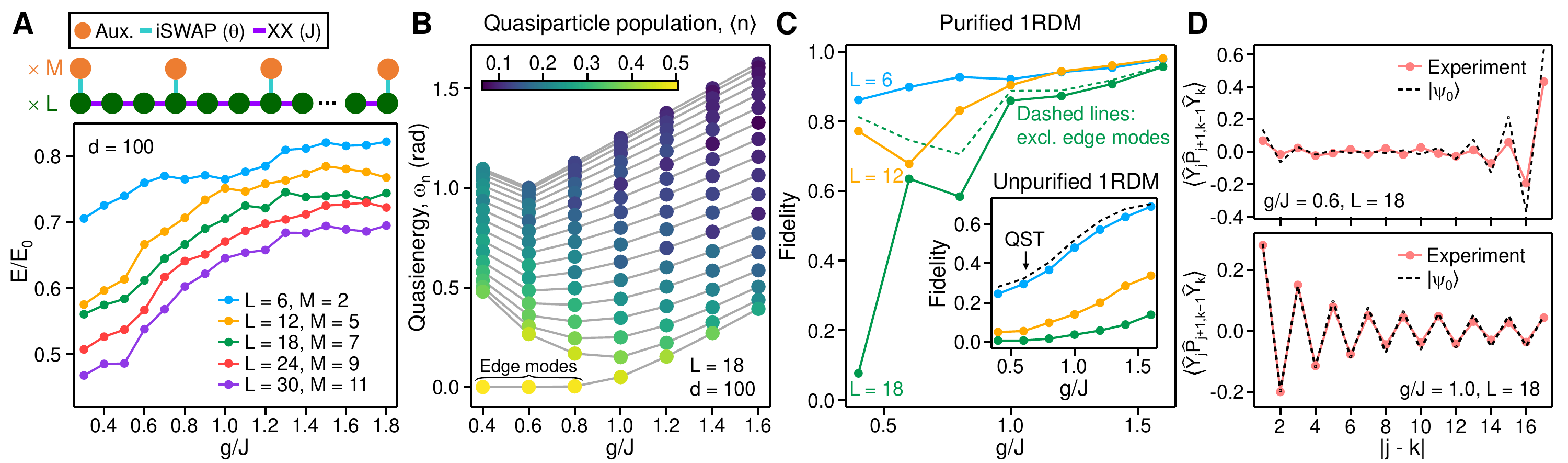} 
    \caption{{Observation of long-range quantum correlations.} (A) Steady state ($d = 100$) energy ratio $E/E_0$ for a TFIM chain with system sizes up to $L = 30$ qubits, plotted as a function of $g/J$. The number of auxiliaries, $M$, is proportionally increased with $L$. (B) Quasienergies $\omega_{n}$ of the $L$ non-interacting quasiparticles describing $\hat{U}$, as a function of $g/J$. The color of each point corresponds to an experimentally measured quasiparticle population $\braket{n}$ at $d = 100$. Here $L = 18$. (C) Ground state fidelities as functions of $g/J$ for system sizes $L = 6$, 12 and 18, constructed from the purified 1RDM. The dashed line refers to $L = 18$ fidelities computed excluding contribution from the edge modes. Inset shows the fidelities without purification. The values obtained from QST in the $L = 6$ case (Fig.~\ref{fig:2}C) are also included and show close agreement with 1RDM estimates for the same system size. (D) Long-ranged correlator $\braket{\hat{Y}_j \hat{P}_{j + 1, k - 1} \hat{Y}_k}$ as a function of $|j - k|$ for $g/J = 0.6$ (upper panel) and $g/J = 1.0$ (lower panel). Results are constructed from the purified 1RDM and exact calculations for the ground state are also shown for comparison.}
    \label{fig:3}
\end{figure*}

To further understand the structure of the steady states, we perform quantum state tomography (QST) and obtain the density matrices $\rho$ of the 6-qubit system at $d = 100$. The results are then used to compute the fidelities with respect to the ground state $\ket{\psi_0}$ and the first excited state $\ket{\psi_1}$ of the TFIM (Fig.~\ref{fig:2}C). We find that $\bra{\psi_0} \rho \ket{\psi_0}$ and $\bra{\psi_1} \rho \ket{\psi_1}$ assume an approximately equal value of 0.26 deep in the antiferromagnetic phase $g/J = 0.4$, indicating that the system cools to an equal mixture of the two nearly degenerate states. As the system approaches the paramagnetic phase, the degeneracy is lifted and $\bra{\psi_0} \rho \ket{\psi_0}$ increases to 0.61 while $\bra{\psi_1} \rho \ket{\psi_1}$ decreases to 0.05 at $g/J = 1.6$. Alternatively, the state fidelities are also computed by choosing $\bra{\psi_0}$ ($\bra{\psi_1}$) to be the ``ground'' (``first excited'') eigenstate of the cycle unitary $\hat{U}$, defined as the state having 0 (1) low-energy quasiparticle excitation (see later discussion). The resulting values are found to be higher, due to the fact that the digital cooling process here is fundamentally governed by time-periodic (a.k.a. Floquet) dynamics rather than a time-independent Hamiltonian. The fixed points of the dissipative evolution are therefore closer to the Floquet eigenstates of the cycle unitary than those of $\hat{H}_\text{TFIM}$. A detailed theoretical treatment of Floquet cooling is presented in the SM.

A key advantage of dissipatively prepared quantum states is that their lifetime extends beyond the coherence times of physical qubits. To test this, we measure the Floquet ground state fidelity $\bra{\psi_0} \rho \ket{\psi_0}$ up to $d = 384$, corresponding to a time duration of 49 $\mu$s ($\gg$ single-qubit $T_1 = 22$ $\mu$s). As shown in Fig.~\ref{fig:2}D, the fidelity exhibits no sign of decay up to this time scale. Furthermore, we find that the nearest-neighbor concurrence increases from 0 to a steady-state value of 0.1 over time, indicating the generation and preservation of entanglement by the cooling process \cite{Wootters_PRL_1998}.

We now test the scalability of the dissipative cooling protocol by extending it to larger system sizes in 1D. A natural starting point is measuring the steady state ($d = 100$) energy with respect to $\hat{H}_\text{TFIM}$, $E$, as a function of $g/J$ (Fig.~\ref{fig:3}A). Here the number of auxiliaries is increased for longer qubit chains to overcome the higher decoherence rates. For an auxiliary-to-qubit ratio of $\frac{M}{L} \approx 0.4$, we observe only weak degradation in $E/E_0$ as the system scales up. In particular, $E/E_0$ retains a value of 0.65 for $L = 30$ compared to a value of 0.76 for $L = 6$ at the critical point $g/J = 1.0$. This result suggests that our protocol is capable of maintaining a low energy density for large 1D quantum systems.

For practical applications of the steady state, it is desirable to improve its fidelity via error-mitigation. One such strategy is purification, which projects a mixed-state experimental density matrix to the closest pure state before computing observables \cite{HF_paper_2020}. Although this is generally challenging for non-integrable quantum systems due to the $O(e^L)$ measurement overhead of full QST, the integrability of the 1D TFIM renders an efficient description of the eigenspectrum of $\hat{U}$ in terms of $L$ non-interacting Bogoliubov fermionic quasiparticles. The many-body spectrum of $\hat{U}$ is represented by the filling or emptying of each quasiparticle level, and each many-body eigenstate belongs to a class of Gaussian states. Such states can be fully characterized through the one-particle reduced density matrix (1RDM) of the quasiparticles, which requires measuring only $O(L^2)$ multiqubit operators (see SM for the exact compositions of these operators).

\begin{figure*}[t!]
    \centering
    \includegraphics[width=2\columnwidth]{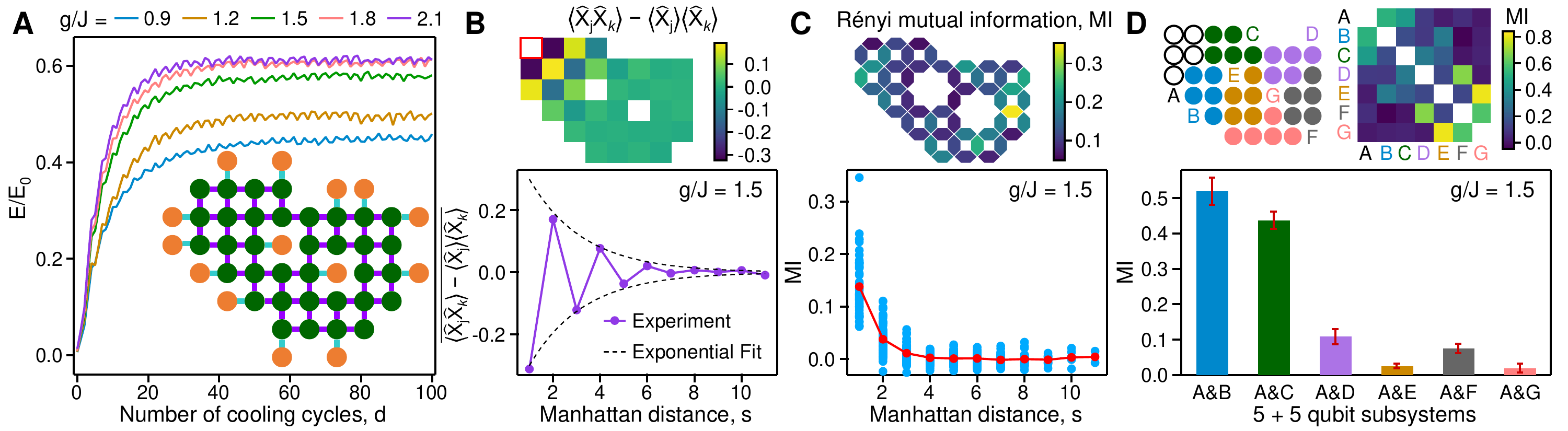} 
    \caption{{Dissipative cooling of a 2D TFIM and many-body mutual information.} (A) $E/E_0$ of a 2D TFIM as a function of $d$ for different values of $g/J$. Inset shows the experimental geometry, where 35 coupled qubits in 2D are connected to 14 auxiliaries. $E_0$ is calculated using the density matrix renormalization group. Here $J$ is chosen to be 0.15 to reduce the Trotter error. (B) Upper panel: Correlator $\braket{\hat{X}_j\hat{X}_k} - \braket{\hat{X}_j} \braket{\hat{X}_k}$ between the top-left qubit (red square) and every other qubit in the system, at the critical point $g/J = 1.5$. Lower panel: $\overline{\braket{\hat{X}_j\hat{X}_k} - \braket{\hat{X}_j} \braket{\hat{X}_k}}$ as a function of the Manhattan distance $s$ between qubits, obtained from the upper panel. The overline denotes averaging over data with the same $s$. Dashed lines show an exponential fit $\sim$$e^{-\frac{s}{2.3}}$ to the decaying envelope. (C) Upper panel: R\'enyi mutual information (MI) between all nearest-neighbor qubits. Lower panel: MI between all possible pairs of qubits in the system (blue) as a function of $s$. Red symbols indicates the average values. (D) Upper panel: MI between different partitions of the system, each containing 5 qubits. Lower: MI between subsystem A and every other subsystem. Error bars indicate standard errors estimated from jacknife resampling.}
    \label{fig:4}
\end{figure*}

Using experimental 1RDMs, we first construct the steady state population $\braket{n}$ for each quasiparticle level across the two different phases. In Fig.~\ref{fig:3}B, the numerically computed quasienergy $\omega_n$ is shown as a function of $g/J$, where the colors of the data points represent measured values of $\braket{n}$. Here the ground state $\ket{\psi_0}$ of the system corresponds to a state where $\braket{n} = 0$ for all quasiparticle levels whereas a trivial depolarized state yields $\braket{n} = 0.5$ for all levels. In comparison, we find that the experimental populations follow a clear distribution whereby $\braket{n}$ increases as $\omega_n$ decreases. In particular, $\braket{n} \approx 0.5$ for the levels associated with the localized edge modes in the antiferromagnetic phase ($g/J < 1$) \cite{Yates_PRB_2019, mi_mem_science_2022}. This dependence is theoretically understood to be a result of the optimal auxiliary energy splitting matching the upper quasiparticle band edge and thus being detuned from the lower band edge (see SM). The quasiparticle populations provide an error budget to the overall cooling performance from each quasiparticle level and allow the ground state fidelity to be estimated, since the probability of finding the system in $\ket{\psi_0}$ is $\prod_{n=1}^L (1 - \braket{n})$.

We then numerically perform a purification of the 1RDMs using a method akin to McWeeny purification used in e.g. quantum chemistry \cite{McWeeny_1960} (see SM). The ground state fidelities, constructed from the quasiparticle populations $\braket{n}$ after purification, are shown in Fig.~\ref{fig:3}C. At the critical point $g/J = 1.0$, we observe fidelities of 0.92 ($L = 6$), 0.90 ($L = 12$) and 0.86 ($L = 18$), which degrade only weakly over system size. In contrast, the fidelities from the unpurified 1RDMs decay exponentially over $L$, as shown in the inset of Fig.~\ref{fig:3}C. The dramatic increase of fidelity through the purification process indicates that despite its mixed nature, the steady state has a large overlap with the ground state of the TFIM in its dominant eigenvector. We note that the fidelity decrease in the antiferromagnetic regime is due to the $\sim$0.5 populations of the edge modes which lead to high purification uncertainties. This interpretation is confirmed by the $L = 18$ fidelities calculated without the edge modes, where the degradation is much reduced (Fig.~\ref{fig:3}C).

Having error-mitigated the steady state, we now demonstrate its topological and quantum-critical behaviors through measuring the long-ranged correlator (Fig.~\ref{fig:3}D):
\begin{equation}
\braket{\hat{C}_{jk}} = \braket{\hat{Y}_j \hat{P}_{j + 1, k - 1} \hat{Y}_k},
\end{equation}
where the parity operator $\hat{P}_{j + 1, k - 1} = \prod_{n=j + 1}^{k - 1} \hat{Z}_n$. In the Majorana-fermion formulation of the 1D TFIM, $\braket{\hat{C}_{jk}}$ is the correlation between Majorana operators on sites $j,k$ of the chain (see SM). We first show $\braket{\hat{C}_{jk}}$ in the antiferromagnetic regime $g/J = 0.6$ (upper panel of Fig.~\ref{fig:3}D). Here we observe that $\braket{\hat{C}_{jk}}$ is nearly zero at short range ($|j - k| \leq 12$) but suddenly increases for $|j - k| > 12$. This is a manifestation of the correlation between the exponentially localized edge modes at the ends of the open chain, which map onto the topological phase of a Majorana chain~\cite{Yates_PRB_2019, mi_mem_science_2022}. At $g/J = 1.0$ (lower panel of Fig.~\ref{fig:3}D), we observe that $\braket{\hat{C}_{jk}}$ has a maximum at $|j - k| = 1$ instead and decays as a power law over distance, consistent with the critical behavior of the TFIM and in close agreement with ground state calculations. In the SM, we also present $\braket{\hat{C}_{jk}}$ in the paramagnetic regime where it decays exponentially due to the ground state resembling a product state (Fig.~S6), and entanglement entropy measurements showing logarithmic growth at the critical point (Fig.~S7 and Fig.~S11).

Although we have thus far focused on exactly solvable models to develop physical insights into dissipative cooling, the experimental protocol is also applicable to non-integrable models where the ground states are not known {\it a priori}. Figure~\ref{fig:4}A shows the dissipative cooling of a 2D TFIM, implemented with 35 qubits connected to 14 auxiliaries. Besides its large size, the 2D model is challenging to cool since each application of $\hat{U}$ includes four layers of parallel two-qubit $XX(J)$ gates, compared to only two layers in 1D. Nevertheless, we find that the system still stabilizes to a low-energy state of the 2D TFIM from a scrambled initial state, with a steady-state energy ratio $E/E_0 = 0.58$ at the critical point ($g/J = 1.5$ for this particular geometry). The antiferromagnetic behavior of the steady state is visible through measurements of the connected correlator $\braket{\hat{X}_j \hat{X}_k} - \braket{\hat{X}_j} \braket{\hat{X}_k}$ between a corner qubit and every other qubit (Fig.~\ref{fig:4}B). We observe that the correlation persists over a Manhattan distance of $\sim$6 sites, with a characteristic decay length of $\sim$2.3.

\begin{figure*}[t!]
    \centering
    \includegraphics[width=2\columnwidth]{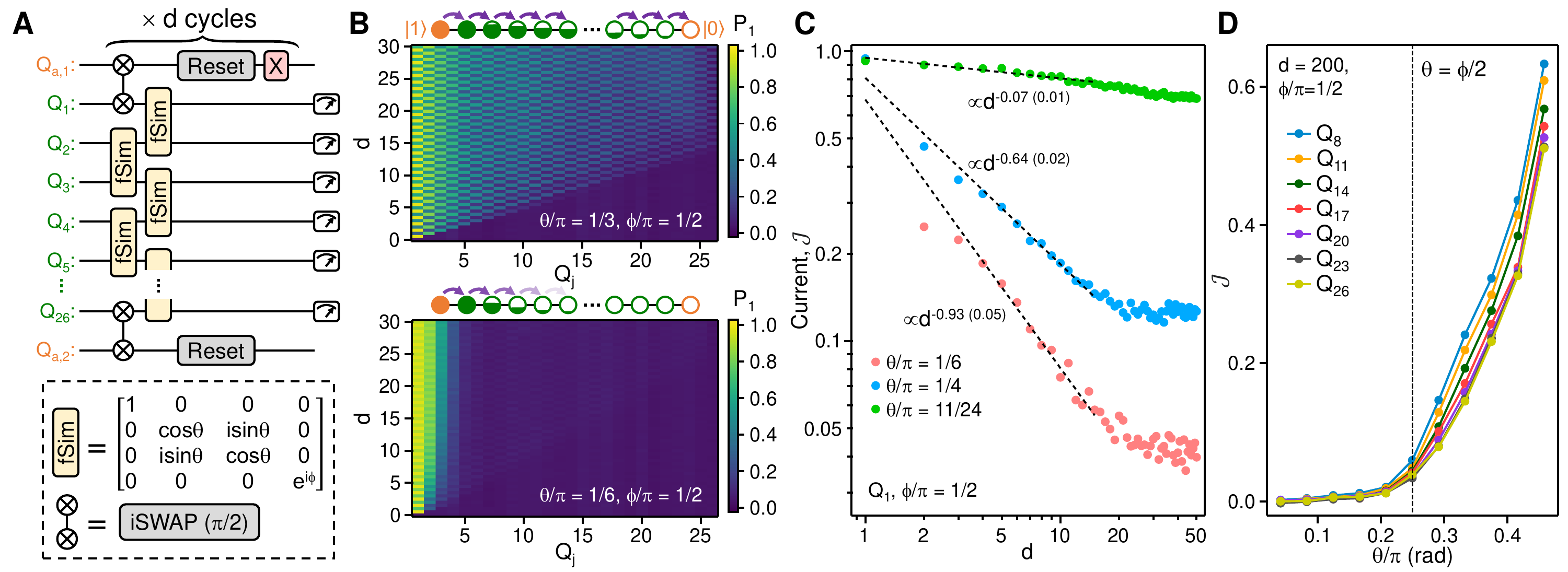} 
    \caption{{Non-equilibrium transport driven by different reservoirs.} (A) Quantum circuit for realizing a boundary-driven Floquet XXZ model. Here a 1D chain of 26 qubits are driven by fSim gates and connected to two auxiliaries on the edges that are reset after every cycle. An $X$ gate is applied to $Q_{\rm a, 1}$ after each reset to stabilize it in the $\ket{1}$ state. (B) $\ket{1}$ state population $P_1$ for different qubits $Q_j$ as a function of $d$, measured with $\phi/\pi = 1/2$ and two different values of $\theta$. Mid-cycle (i.e. $d + 0.5$) data for $P_1$ are also included, which are taken between the two layers of fSim gates within each cycle. (C) Current flow $\mathcal{J}$ from $Q_{\rm a, 1}$ to $Q_1$, extracted via $\mathcal{J}(d) = P_1 (d + 0.5) - P_1 (d)$ where $P_1$ is measured at $Q_1$, as a function of $d$. Dashed lines are fits to early-time ($d \leq 15$) data using the function form $A d^\alpha$, where $A$ and $\alpha$ are free parameters. The fitted results for $\alpha$ are shown as exponents of $d$, along with their standard errors in parentheses.  (D) Steady state ($d = 200$) $\mathcal{J}$ as a function of $\theta$, shown for different qubits along the chain. $\mathcal{J}$ represents the population transfer from $Q_{j - 1}$ to $Q_j$ in each cycle and is extracted by $\mathcal{J}(d) = P_1 (d + 0.5) - P_1 (d)$ for $Q_j$ with odd $j$ and $\mathcal{J}(d) = P_1 (d + 1) - P_1 (d + 0.5)$ for $Q_j$ with even $j$. Dashed line indicates the isotropic point $\theta = \phi/2$. }
    \label{fig:5}
\end{figure*}

To probe the entanglement structure of the steady state, we adopt the second-order R\'enyi mutual information:
\begin{equation}
\text{MI} = S^{(2)}_\text{A} + S^{(2)}_\text{B} - S^{(2)}_\text{AB},
\end{equation}
where $S^{(2)} = -\log_2 \text{Tr}{\rho^ 2}$ denotes the second-order R\'enyi entropy of a subsystem (A, B or AB) with density matrix $\rho$. MI includes contributions from both classical and quantum correlations and is relatively insensitive to classical entropy coming from imperfect cooling or measurement errors \cite{Islam2015}. In 2D, the MI is generally inaccessible to quantum simulators in which measurements are limited to a single basis \cite{Mazurenko2017, Ebadi2021}. Leveraging the universal gate set of the quantum processor, we obtain MI between all possible partitions of the system through a single set of randomized measurements \cite{Brydges_2019} (see SM).

The upper panel of Fig.~\ref{fig:4}C shows the MI between nearest-neighbor qubits, where values between 0.06 and 0.35 are observed throughout the system. The MI between all qubit pairs is shown in the lower panel of Fig.~\ref{fig:4}C, where it is seen to decay over distance. Despite the spatial decay, MI is finite between qubits separated by $s \approx 3$. We note that the fluctuation of MIs between qubits separated by the same Manhattan distance is likely due to inhomogeneous cooling across the system due to, e.g., different qubit decoherence rates. The randomized measurements also allow us to extract the many-body MI between seven 5-qubit partitions of the system, as shown in the upper panel of Fig.~\ref{fig:4}D. Here we again observe a large MI between contiguous 5-qubit subsystems, which decays as the subsystems become more separated (lower panel of Fig.~\ref{fig:4}D). Notably, we still observe finite MIs between some non-neighboring subsystems, such as A and D. The behavior of MI above shows that the cooling protocol is capable of steering models of quantum magnetism into correlated steady states. Further improvements of qubit coherence times will allow preparation of a large variety of magnetic states with longer-ranged quantum correlations.

The dissipative dynamics investigated thus far has focused on coupling a many-body system to a single reservoir. It is natural to ask whether quantum-coherent behavior may also arise from coupling the system to different reservoirs, which induces non-equilibrium transport through a chemical potential difference. We explore this possibility using another paradigmatic model of quantum magnetism, the 1D XXZ spin chain \cite{LIEB1961407}, which is currently the subject of intense theoretical \cite{BertiniRMP2021} and experimental \cite{scheie2021detection,BlochKPZ2021,Keenan2023} investigations due to its rich magnetic transport properties. A Floquet version of the XXZ model~\cite{Gritsev17integrable, morvan2022formation} is readily implementable using consecutive applications of fSim gates parameterized by a conditional phase $\phi$ and iSWAP angle $\theta$, as shown in Fig.~\ref{fig:5}A. Here the qubit $\ket{0}$ ($\ket{1}$) state mimics the spin-up (spin-down) state. A pair of boundary auxiliaries ($Q_{\rm a, 1}$ and $Q_{\rm a, 2}$), stabilized to $\ket{1}$ and $\ket{0}$ states, are coupled to the chain via iSWAP gates. We then measure $\ket{1}$ state probability, $P_1$, of the system qubits (initialized in $\ket{0}^{\otimes L}$) over $d$.  

Within the linear response regime ~\cite{ZotosPRB1996, ZnidaricPRL2011}, the XXZ chain is predicted to show different transport regimes depending on the anisotropy parameter $\Delta=\frac{\phi}{2\theta}$. In our experiment, the strong driving from the boundary auxiliaries unveils transport phenomena in a highly non-equilibrium regime far away from linear response. The initial spreading of qubit excitations in the system up to $d = 30$ is illustrated in Fig.~\ref{fig:5}B. In the easy-plane regime ($\Delta < 1$), we observe a ballistic propagation consistent with the existence of freely propagating magnon quasiparticles. In contrast, in the easy-axis ($\Delta > 1$) regime, qubit excitations fail to propagate into the system. Instead, a relatively sharp domain wall is formed between a few excited qubits adjacent to ${Q}_{1}$ and the other qubits which remain in the $\ket{0}$ state. The observed domain wall is due to the fact that $n$ adjacent qubit excitations form a heavy bound state with a group velocity exponentially suppressed by $n$, hence inhibiting transport \cite{morvan2022formation}.

We next characterize details of the quantum transport through the local current $\mathcal{J}$, which corresponds to the difference between qubit populations in the middle and at the end of each cycle $d$. Compared to $P_1$, $\mathcal{J}$ is less sensitive to readout errors. Figure~\ref{fig:5}C shows the time-dependent $\mathcal{J}$ at $Q_1$, corresponding to the population pumped into the system from $Q_{\rm a, 1}$ per cycle \cite{note}. At early cycles ($d \leq 15$) where qubit decoherence plays a minor role in transport, we observe different dynamical exponents depending on $\Delta$: In the easy-plane regime, the current is nearly constant at early times and scales as $\propto d ^ {-0.07}$. In the easy-axis regime, we find a dependence $\mathcal{J}(d)\propto d ^ {-0.93}$, which corresponds to a total population transfer approximately scaling as $\sim$$\log d$. The unusual logarithmic scaling was found in a recent Bethe ansatz solution for the Hamiltonian case \cite{buvca2020bethe}. At the isotropic point ($\Delta = 1$) where no exact solution is available, we observe a power-law scaling $\mathcal{J}(d)\propto d^{a}$ with a sub-diffusive exponent $a \approx -0.64$. The dynamical exponent found at the isotropic point agrees well with noise-free numerical simulation of larger system sizes shown in the SM, which finds a similar value of $a \approx -0.72$. This result qualitatively differs from recent experiments in closed quantum systems which observed super-diffusive ($a = -1/3$) transport at the isotropic point \cite{scheie2021detection, BlochKPZ2021}, providing evidence for a previously unknown transport regime of the XXZ model outside linear response.

Lastly, we focus on the long-time behavior of the local currents after their saturation at $d\gtrsim 30$. The saturation corresponds to the formation of a non-equilibrium steady state (NESS) which is stable up to an experimental limit of $d = 200$. Interestingly, despite the spreading of qubit decoherence at this late cycle, we observe that $\mathcal{J}$ still depends sharply on the coupling anisotropy and serves as a dynamical order parameter for the different transport regimes (Fig.~\ref{fig:5}D): In the easy-plane regime, we observe finite current flow through qubits away from $Q_{\rm a,1}$. At the isotropic point, $\mathcal{J}$ is greatly suppressed but retains a finite value throughout the chain. In the easy-axis regime, we find that $\mathcal{J}$ is nearly 0 for all qubits. Our results complement early theoretical investigations of a boundary-driven XXZ model and indicate that the insulating behavior of the XXZ model persists even in the presence of qubit decoherence \cite{ProPRL11}.

In summary, our work highlights engineered dissipation as a promising method for preparing quantum many-body states. Compared to state preparation algorithms based on unitary evolution, our protocol has several advantages, including long lifetimes of the prepared states, robustness across quantum phase transitions and a better scaling at larger system sizes (see Section S4 of the SM). Even against variational quantum algorithms which may achieve similar performance at current system sizes, the dissipative protocol is advantageous owing to its minimal optimization overhead and the ability to capture long-range quantum correlation. Despite these advantages, we note that cooling generic Hamiltonians may require complex system-bath couplings that are too challenging to implement in practice. The development of error-mitigation schemes for non-integrable models such as the 2D TFIM remains another open question.

Beyond cooling, we find that our platform may also be applied to study non-equilibrium dynamics that is difficult to access via closed quantum systems. Using the XXZ chain as an example, we have already made the experimental discovery of a new transport regime in this well-known quantum spin model. Our work therefore broadly enhances the functionality of quantum processors by introducing engineered dissipative channels as fundamental building blocks, with applications to open-system quantum simulation \cite{Kapit_PRR_2020}, quantum transport \cite{Harris2017} and stabilization of topological quantum states \cite{Diehl2011,Bardyn_2013, Kapit_PRX_2014}.

{\bf Acknowledgements --} We acknowledge fruitful discussions with M.~H.~Devoret, R.~Moessner, E.~Kapit and A.~Blais.

{\bf Author Contributions --} X.~M. , V.~Smelyanskiy and D.~A.~A. conceived and designed the experiment. X.~M. performed the experiment. V.~Smelyanskiy and D.~A.~A. developed the theory of quasiparticle and Floquet cooling. A.~A.~M. performed numerical simulations and analyzed the data in the manuscript. S.~S. , J.~Lloyd and E.~R. performed additional numerics. K.~C.~M. developed the fast qubit reset. P.~V.~K. developed gate optimization routines. X.~M. , A.~A.~M. , V.~Smelyanskiy, D.~A.~A. and P.~R. wrote the manuscript. Infrastructure support was provided by Google Quantum AI. All authors contributed to revising the manuscript and the Supplementary Materials.

{\bf Data availability --} Xiao Mi, Data for ``Stable Quantum-Correlated Many Body States via Engineered Dissipation,'' Zenodo (2023); \href{https://doi.org/10.5281/zenodo.8187929}{https://doi.org/10.5281/zenodo.8187929}.

\twocolumngrid
\bibliographystyle{Science}
\bibliography{cooling.bib}

\begin{thebibliography}{10}
\bibitem{note}
The current from $Q_{26}$ to $Q_{\rm a, 2}$ should in principle be subtracted from $J (Q_1)$ to account for the particle loss into the drain. Since we are focusing on early time ($d < 15$) dynamics, this current is near or exactly 0 and does not affect the transport exponent. See SM for an explicit verification.
\end{thebibliography}


\newpage
\onecolumngrid




\setcounter{equation}{0}
\setcounter{figure}{0}
\setcounter{table}{0}
\setcounter{page}{1}
\makeatletter
\renewcommand{\thesection}{S\arabic{section}}
\renewcommand{\theequation}{S\arabic{equation}}
\renewcommand{\thefigure}{S\arabic{figure}}
\renewcommand{\thetable}{S\arabic{table}}

\newpage

\begin{center}
    \textbf{\large Supplementary Materials for ``Stable Quantum-Correlated Many Body States through Engineered Dissipation''}
\end{center}
\vspace{18pt}

\section{Experimental details and additional data}

\subsection{CPHASE and fSim gates}

The quantum processor used in our experiment is similar to those used in recent works \cite{mi_mem_science_2022,Acharya2023} and consists of 68 frequency-tunable transmons with tunable couplings. The median $T_1$ of the qubits is 22 $\mu$s. Two main improvements have been made to the calibration of tunable CPHASE gates and tunable fSim gates to enable dissipative cooling and XXZ non-equilibrium transport: (i) The errors of the CPHASE gates are reduced by more than two-fold compared to past implementations \cite{Mi_DTC_2022}. (ii) The tunability of the fSim gates is enhanced \cite{morvan2022formation}, allowing nearly all combinations of iSWAP angles and conditional phases to be accessed with low control errors. We briefly outline the technical progress that enabled these improvements below.

The CPHASE gates are implemented by flux pulses that bring two transmons to a relative frequency detuning of $\epsilon_\text{2p}$ between the $\ket{11}$ and $\ket{02}$ states, while ramping the coupler to enact a $XX + YY$ coupling with a maximum value of $g_\text{max}$. After a pulse duration $t_\text{p} \approx \frac{1}{\sqrt{8 g_\text{max}^2 + \epsilon_\text{2p}^2}}$, qubit leakage (i.e. $\ket{2}$ state population) returns to $\sim$0 and a conditional phase $\phi$ is accumulated. By keeping $t_\text{p}$ fixed while adjusting $g_\text{max}$ and $\epsilon$, $\phi$ may be tuned between $0$ and $2 \pi$. The gate fidelities were found to be $\sim$$99.0 \%$ in our earlier works \cite{Mi_DTC_2022}, limited by both coherence times and a parasitic iSWAP angle $\theta \approx 0.02$ rad.

In the current work, we have reduced the parasitic iSWAP angles to a lower level of $\sim$0.003 rad. This is accomplished by smoothing the coupler pulses to better maintain an adiabatic condition between the $\ket{01}$ and $\ket{10}$ states of the qubits during gate implementation. The reduced iSWAP errors also allow us to remove the qubit detuning pulses (a.k.a. ``physical'' $Z$ gates) before the coupler pulses and replace them with virtual $Z$ rotations, which further reduce gate errors \cite{Mi_OTOC_2021}. Furthermore, improved frequency-selection algorithms along with careful monitoring of system stability allow avoidance of two-level system (TLS) defects in large quantum systems, reducing the number of outlier gates \cite{Klimov_PRL_2018, Klimov_Snake_2020}. Lastly, we have also optimized the fidelity of single-qubit gates by reducing their pulse lengths.

\begin{figure}[h]
    \centering
    \includegraphics[width=0.5\columnwidth]{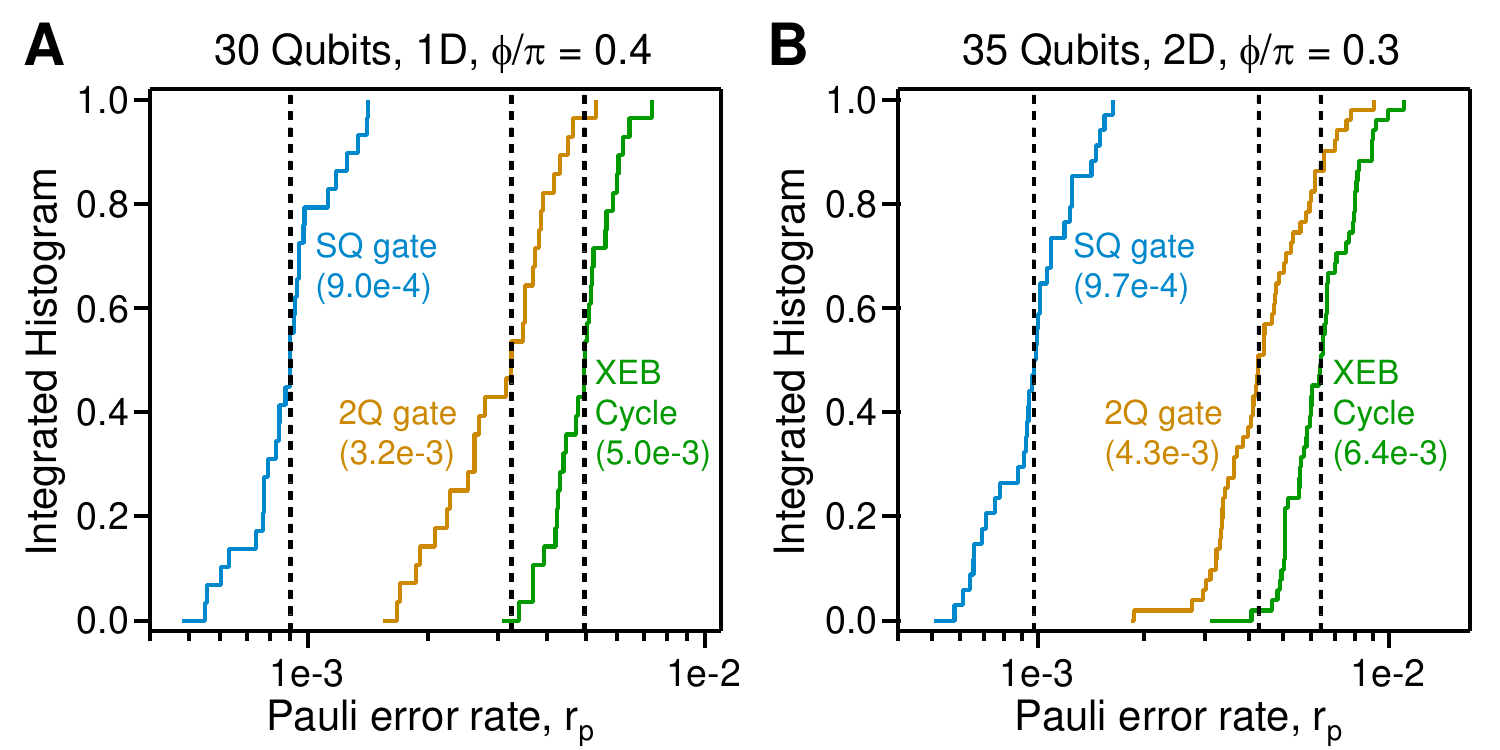} 
    \caption{{Single-qubit and CPHASE gate fidelities.} (A) Integrated histograms of Pauli error rates $r_\text{p}$ associated with single-qubit $\sqrt{X}$ and $\sqrt{Y}$ rotations (blue), two-qubit CPHASE gate with $\phi / \pi = 0.4$ (brown) and an XEB cycle (green). Median value of each histogram is listed within the figure and also indicated with a vertical dashed line. The results are obtained with a 1D chain of 30 qubits and gates executed in parallel. (B) Same as panel {\bf a} but with $\phi / \pi = 0.3$ and 35 qubits in 2D. Gates are also executed in parallel.}
    \label{fig:s1}
\end{figure}

\begin{figure}[t!]
    \centering
    \includegraphics[width=1\columnwidth]{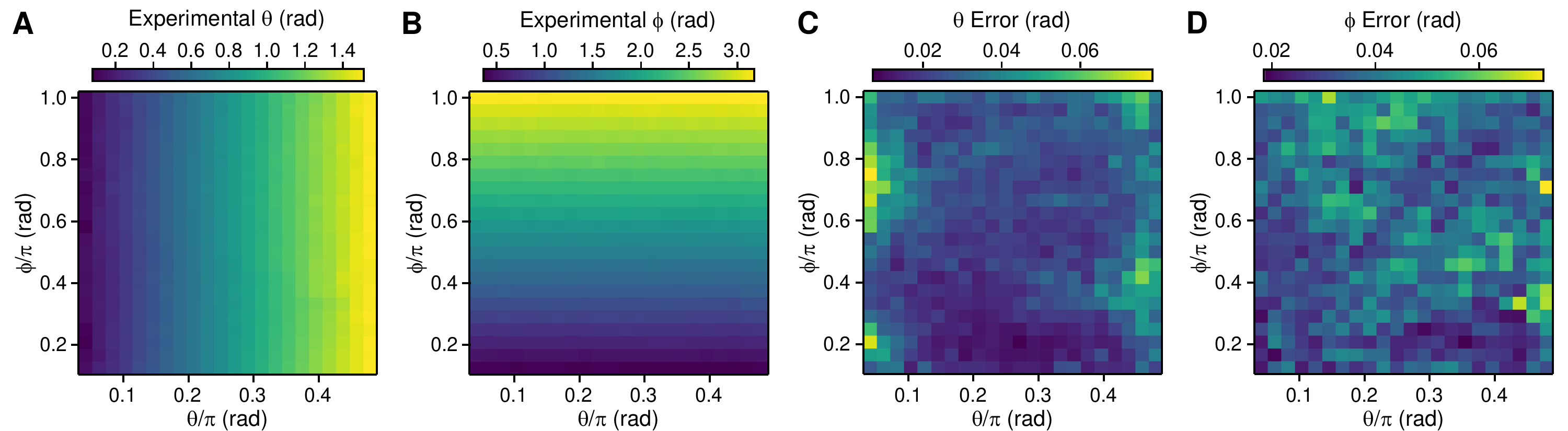}
    \caption{{Continuously tunable fSim gates.} (A) The experimentally measured values of $\theta$ as a function of target $\theta$ and $\phi$, averaged over 25 qubit pairs in a chain of 26 qubits. (B) Same as panel A but with experimentally measured values of $\phi$ plotted. (C) Root-mean-squared differences between target and measured values of $\theta$, averaged over all 25 qubit pairs. (D) Same as panel C but with the error in $\phi$ shown.}
    \label{fig:s2}
\end{figure}

The fidelities of single-qubit gates and two-qubit CPHASE gates are shown in Fig.~\ref{fig:s1}. Here we show the gate errors associated with the 30-qubit 1D chain used in Fig.~3A and the 35-qubit 2D grid used in Fig.~4 of the main text, along with the conditional phases $\phi$ used in these two figures. In 1D, we achieve single-qubit gate errors that have both a median and a mean of $9.0 \times 10^{-4}$, characterized through simultaneous randomized benchmarking. The two-qubit CPHASE gate errors, characterized through simultaneous cross-entropy benchmarking \cite{Arute2019}, have a median (mean) error of $3.2 \times 10^{-3}$ ($3.1 \times 10^{-3}$). In 2D, the single-qubit gate errors have a median (mean) value of $9.7 \times 10^{-3}$ ($9.9 \times 10^{-3}$). The two-qubit CPHASE gate errors have a median (mean) error of $4.3 \times 10^{-3}$ ($4.6 \times 10^{-3}$). For both 1D and 2D, the two-qubit XEB cycle errors (which include contributions from two single-qubit gates and one CPHASE gate) are also included for reference. The entangling gate fidelity is among the lowest for experimentally reported quantum processors of this size.

In contrast to the CPHASE gates, the tunable fSim gates are implemented primarily through the resonant interaction between the $\ket{01}$ and $\ket{10}$ states of the two qubits \cite{morvan2022formation}. In our past works, independent tuning of $\phi$ and $\theta$ is achieved by exploiting the different scaling of each angle with respect to the strength of the interqubit coupling $g_\text{max}$, i.e. $\theta \propto g_\text{max}$ whereas $\phi \propto g_\text{max}^2$. However, achieving full coverage over the entire range of $\phi$ and $\theta$ is difficult, since certain combinations of the two angles require either excessively long pulses or large $g_\text{max}$ where either decoherence or leakage becomes an issue. To circumvent these constraints, we have added a third tuning parameter to the fSim gate, namely a variable detuning $\epsilon_\text{1p}$ between the $\ket{01}$ and $\ket{10}$ states. The detuning parameter allows $\theta$ to be varied while leaving $\phi$ largely constant, allowing coverage of previously unachievable angles.

In Fig.~\ref{fig:s2}A and Fig.~\ref{fig:s2}B, we show experimentally measured values of $\theta$ and $\phi$ over a nearly complete coverage of all possible target angles, $\theta \in [\frac{\pi}{24}, \frac{23\pi}{48}]$ and $\phi \in [\frac{\pi}{8}, \pi]$, averaged over the 26-qubit chain used in Fig.~5 of the main text. The results are obtained from unitary tomography measurements \cite{morvan2022formation}. The control errors associated with each angle are shown in Fig.~\ref{fig:s2}C and Fig.~\ref{fig:s2}D. The average error in $\theta$ is 0.026 rad and the average error in $\phi$ is 0.037 rad. These errors may be reduced in future experiments using Floquet calibration \cite{Neill_Nature_2021}. We also note that the control errors in $\theta$ are larger for $\theta \rightarrow 0$ or $\theta \rightarrow \pi / 2$, which may be a result of their higher sensitivities to state preparation and measurement (SPAM) errors in unitary tomography.

\subsection{Stabilization of single-qubit states}

\begin{figure}[t!]
    \centering
    \includegraphics[width=0.5\columnwidth]{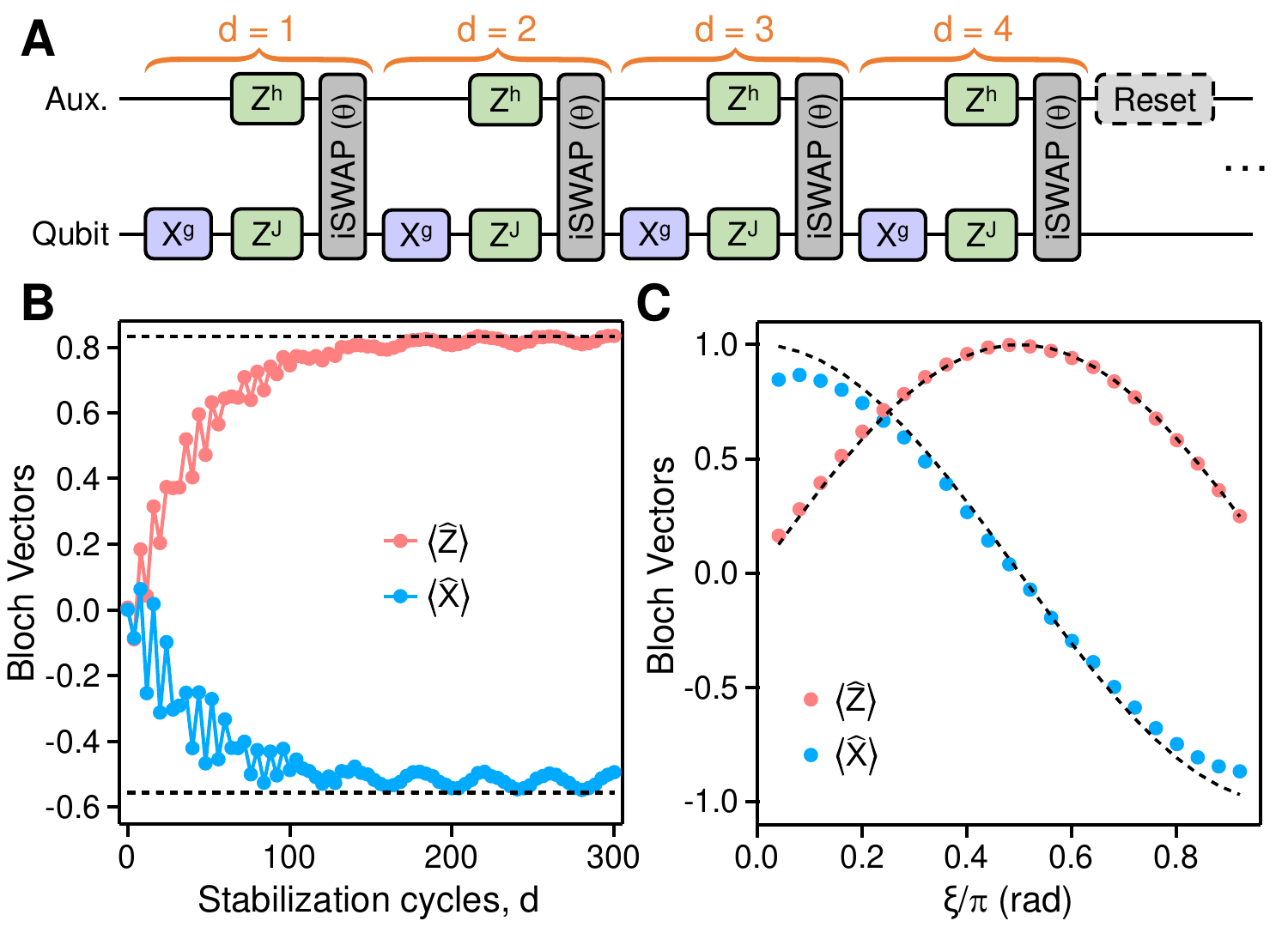}
    \caption{{Stabilization of single-qubit states.} (A) Circuit schematic for stabilizing states of a single qubit using a single auxiliary. (B) Bloch vectors of the single qubit, $\braket{\hat{Z}}$ and $\braket{\hat{X}}$, as a function of number of stabilization cycles $d$. Here $J = 0.18$, $g = -0.12$, $\theta = 0.09$ rad and $h = \sqrt{g^2 + J^2}$. Dashed lines indicate the Bloch vectors corresponding to an eigenstate of the single-qubit Hamiltonian $\hat{H}_{\text 1q} = g \hat{X} + J \hat{Z}$. Readout errors have been corrected in the data via experimentally obtained readout errors. (C) Bloch vectors (averaged between $d = 280$ and $d = 300$) as a function of $\xi$, where $J = A \sin{\xi}$,  $g = A \cos{\xi}$ and $A = \frac{0.3}{|\sin{\xi}| + |\cos{\xi}|}$. $\theta = 0.09$ rad in this plot. Dashed lines indicate the Bloch vectors corresponding to an eigenstate of the single-qubit Hamiltonian at each $\xi$.}
    \label{fig:s3}
\end{figure}

While we have primarily focused on the stabilization of multiqubit systems in the main text, the dissipative scheme is straightforwardly applicable to single-qubit stabilization as well. Past works on this topic have employed superconducting resonators with tailored shot-noise spectrum or parametrically modulated coupling to stabilize the states of a transmon qubit \cite{Murch_PRL_2012,Yao_PRL_2017}. Here we utilize the same setup as Fig.~1 of the main text and seek to stabilize a single qubit to an eigenstate of the Hamiltonian, $\hat{H}_{\text 1q} = g \hat{X} + J \hat{Z}$, by coupling it to a dissipative auxiliary (Fig.~\ref{fig:s3}A). The single qubit is evolved via a Trotterized implementation of $e^{-i \hat{H}_{\text 1q} t}$ using alternating layers of $X^g$ and $Z^J$ gates. The auxiliary is evolved by a phase gate $Z^h$ with an exponent $h = \sqrt{g^2 + J^2}$ that matches the energy splitting of the auxiliary to the qubit. Similar to the TFIM, the qubit and auxiliary are coupled by a weak partial-iSWAP gate having an angle $\theta = 0.09$ rad. A reset is applied to the auxiliary every 4 stabilization cycles, $d$.

The time dependence of the Bloch vectors of the qubit is shown in Fig.~\ref{fig:s3}B, where we have averaged over 20 random initial states. We observe that, on average, the Bloch vectors $\braket{\hat{Z}}$ and $\braket{\hat{X}}$ increase and reach steady state values close to the calculated values for an eigenstate of $\hat{H}_{\text 1q}$. To see how the steady state Bloch vectors compare to the idealized values across different Hamiltonian parameters, we vary the ratio of $g/J$ by sweeping the parameter $\xi = \tan^{-1} {(J/g)}$ and measure the steady state values of the Bloch vectors. The results, plotted in Fig.~\ref{fig:s3}C, show close agreement between the idealized values and the experimentally measured values over a wide range of $\xi$. 

\subsection{Circuit optimization and comparison with quantum trajectory simulations}

\begin{figure}[t!]
    \centering
    \includegraphics[width=1\columnwidth]{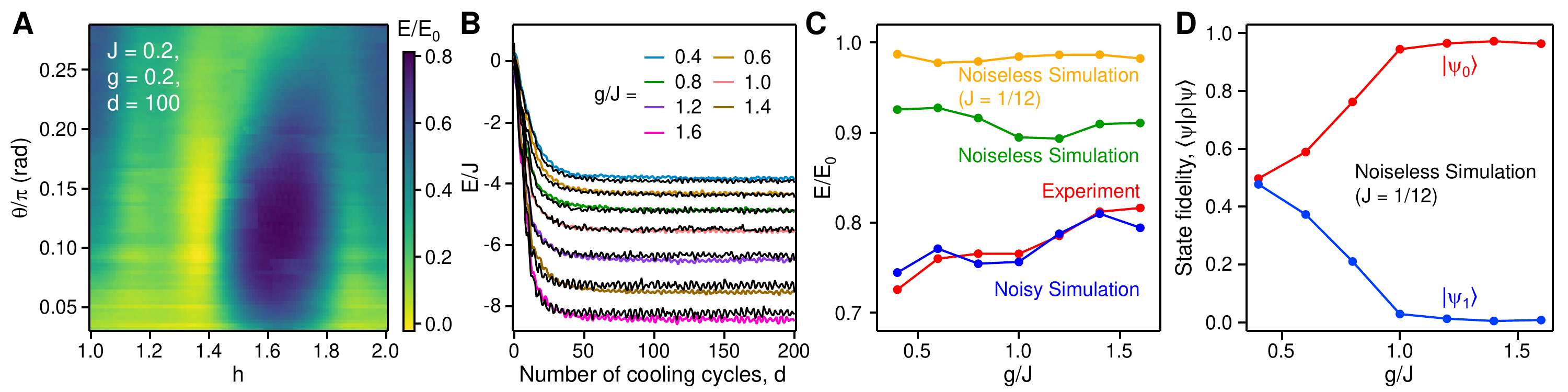}
    \caption{{Circuit parameter optimization and comparison with quantum trajectory simulations.} (A) Experimentally measured ratio between the energy $E$ at $d = 100$ and the ground state energy $E_0$, as a function of gate parameters $\theta$ and $h$ (see Fig.~2 of the main text). $L = 6$ in this plot. (B) Experimentally obtained (colored lines) and numerically simulated (black dashed lines) energies $E$ of the 6-site TFIM as a function of $d$ and for different values of $g/J$. (C) Steady state energy ratio $E/E_0$ as a function of $g/J$, obtained through experiment (red), noisy simulation (blue) and noiseless simulation (green). Noiseless simulation results using a smaller Trotter angle $J = 1/12$ are also shown, where the auxiliary reset is applied every 12 cycles. (D) Steady state fidelities with respect to the ground ($\ket{\psi_0}$) and excited ($\ket{\psi_1}$) states of the 6-site TFIM, computed from noiseless simulation and $J = 1/12$.}
    \label{fig:s4}
\end{figure}

As illustrated by Fig.~1 of the main text, to maximize the efficiency of the dissipative cooling protocol, the energy splitting of the auxiliaries needs to match excitation energies of the quantum system. At the same time, the auxiliary-system coupling needs to be strong enough to remove system excitations at a high rate but weak enough to avoid dressing the energy spectrum of the system and modifying its Hamiltonian. We perform an experimental optimization procedure by measuring the energy of the system $E = \braket{\hat{H}_\text{TFIM}}$ at a late time $d = 100$, while sweeping circuit parameters $\theta$ and $h$. The normalized energy, $E / E_0$, where $E_0$ is the numerically calculated energy of the ground state, is shown in Fig.~\ref{fig:s4}A. A maximum ratio of $E / E_0 \approx 0.8$ is observed at $h = 1.65$ and $\theta / \pi = 0.11$, indicating that the system is closest to the ground state for these circuit parameters. This optimization process is performed for each value of $g/J$ and each different geometry in Fig.~2 to Fig.~4 of the main text.

To confirm that our system is indeed performing as expected, we compare experimentally obtained energies $E$ against numerical simulations of the exact same quantum circuits via quantum trajectory methods. Both experimental results and noisy simulation results are shown in Fig.~\ref{fig:s4}B. Here we have used single-qubit $T_1 = 21$ $\mu$s and $T_2 = 8$ $\mu$s, which are close to the typical coherence times of our qubits. The gate times used in the numerical simulations are also identical to those used in experiments. We find excellent agreements between the numerically simulated time-dependent energies and experimental values, indicating that relatively simple error channels including qubit relaxation and dephasing are sufficient to account for the experimental performance.

Figure~\ref{fig:s4}C shows the steady state energy ratio $E/E_0$ as a function of $g/J$, from both experimental results and noisy simulation. We again observe close agreement between the two cases. To identify the limitation of the cooling performance from decoherence alone, we also simulate the experiment without qubit relaxation and dephasing. The results, also plotted in Fig.~\ref{fig:s4}C, show an improved steady state energy ratio of $E/E_0 \approx 0.9$. The energy ratio in the noiseless simulation is primarily limited by the relatively large Trotter angle, $J = 0.2$ or $J = 0.25$, in these experiments, which is chosen to ensure sufficiently fast motion of quasiparticles such that they are removed by the auxiliaries within a time scale $\ll T_1, T_2$. This is confirmed by reducing the value of $J$ to 1/12 and rerunning the noiseless simulation, the result of which is also shown in Fig.~\ref{fig:s4}C. Here the energy ratio $E/E_0$ is higher and averages to 0.98. As observed in the experiments, the limitation imposed by large $J$ can be mitigated by comparing the steady state to the Floquet eigenstates instead. The fidelities of the steady state with respect to the two low-lying states of the TFIM, for the noiseless simulation with $J = 1/12$, are also shown in Fig.~\ref{fig:s4}D. We observe a total fidelity of $\bra{\psi_0} \rho \ket{\psi_0} + \bra{\psi_1} \rho \ket{\psi_1} > 0.97$ throughout the different phases of the model, with a nearly equal mixture of the two states deep in the anti-ferromagnetic phase ($g/J = 0.4$) owing to the near-degeneracy of the two low-lying states. These results indicate that the cooling protocol can yield high-fidelity results in the idealized case of no decoherence and small Trotter steps.

\begin{figure}[t!]
    \centering
    \includegraphics[width=1\columnwidth]{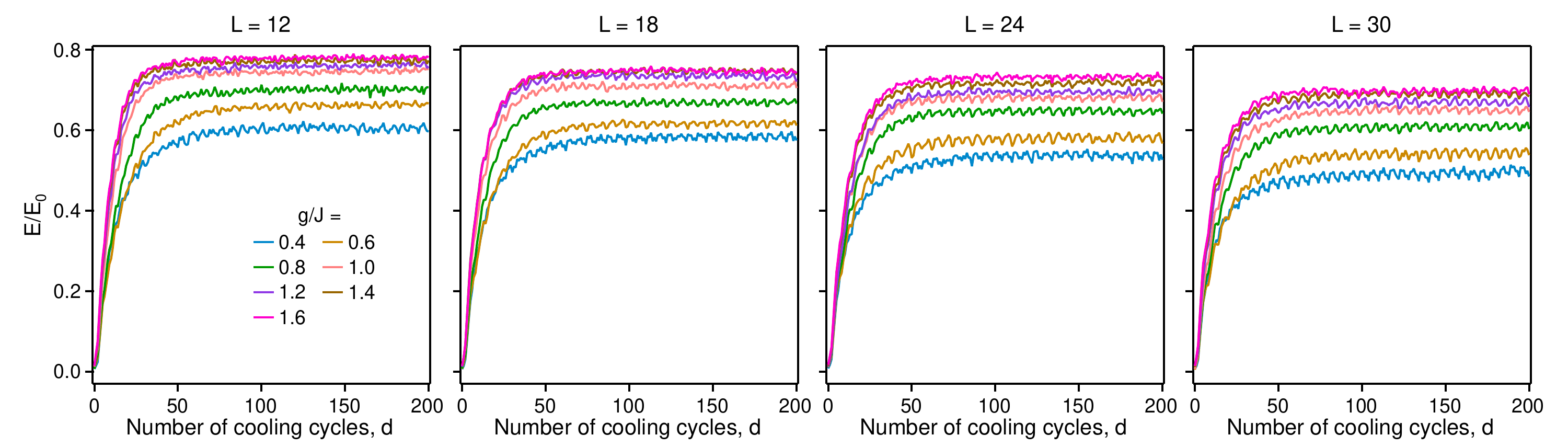}
    \caption{{\bf Time-dependence of energy for large 1D TFIM chains.} Ratio between the measured energy $E = \braket{\hat{H}_\text{TFIM}}$ and the ground state energy $E_0$ as a function of the number of cooling cycles $d$. Data are shown for chain lengths of $L = 12$, 18, 24 and 30.}
    \label{fig:s5}
\end{figure}

\subsection{Time-dependent energy for large 1D TFIM}

The experimental data showing the detailed time dependence of the ratio between measured energy $E$ of the system and the numerically calculated ground state energy $E_0$ of the 1D TFIM is plotted in Fig.~\ref{fig:s5}. Data for each system size $L$ is also shown for different values of $g / J$ spanning the anti-ferromagnetic regime ($g/J < 1.0$), critical point ($g/J = 1.0$) and the paramagnetic regime ($g/J > 1.0$). For each case, we observe that the system reaches a steady state at $d \approx 100$, beyond which $E / E_0$ is approximately constant.

\begin{figure}[t!]
    \centering
    \includegraphics[width=1\columnwidth]{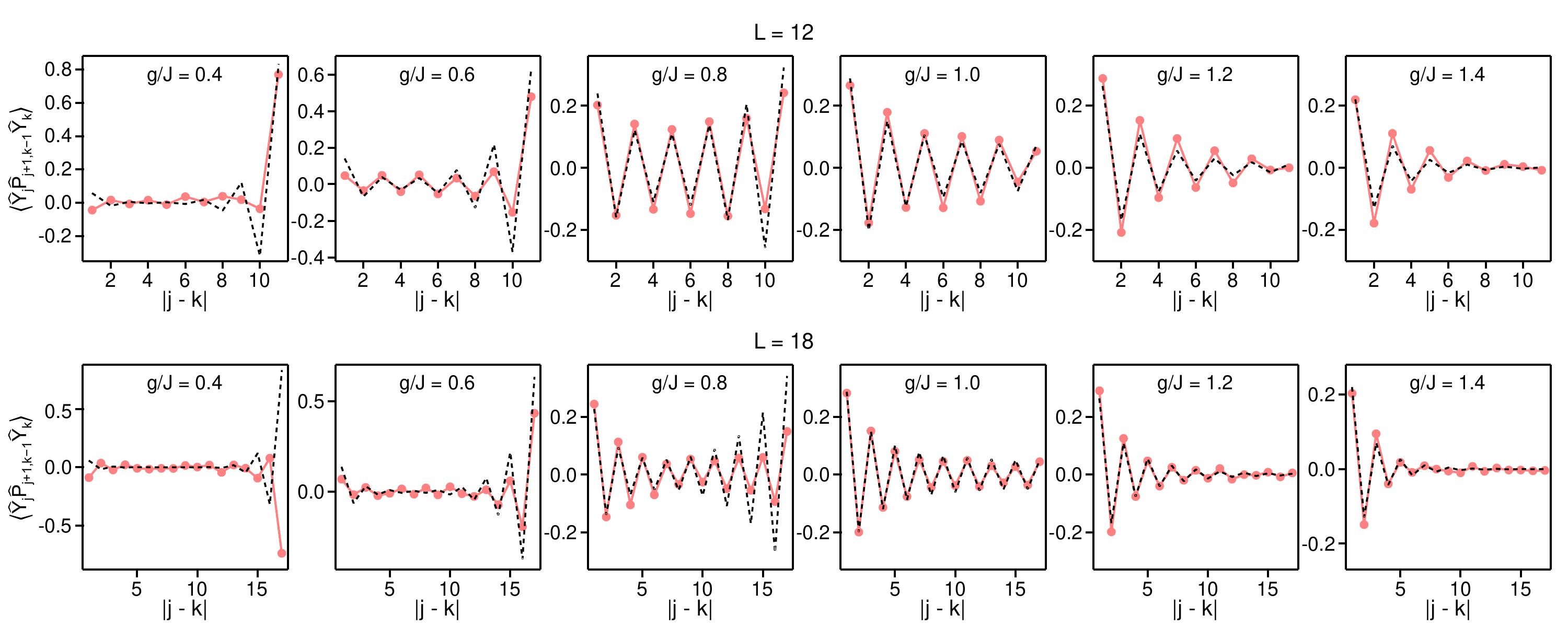}
    \caption{Detailed experimental data of quantum correlations $\braket{\hat{Y}_j \hat{P}_{j + 1, k - 1} \hat{Y}_k}$ for $L = 12$ and $L = 18$, across the quantum phase transition. Solid symbols are experimental results constructed from purified 1RDMs and dashed lines are exact ground state calculations.}
    \label{fig:s6}
\end{figure}

\subsection{Additional quantum correlation data}

Experimentally measured values of the long-range quantum correlator $\braket{\hat{Y}_j \hat{P}_{j + 1, k - 1} \hat{Y}_k}$, constructed using the purified 1RDMs, are shown in Fig.~\ref{fig:s6} for different values of $g/J$ across different phases of the 1D TFIM. Results for both $L = 12$ and $L = 18$. We observe that for $L = 12$, results are in good agreement with the ground state. For $L = 18$, we observe equally good agreement except at $g/J = 0.4$, where the experimental data show oscillations with an opposite sign compared to the ground state, at large $|j - k|$. This is due to the small energy splitting between the two edge modes at this system size and $g/J$, which makes distinguishing them difficult in experiment. Consequently, the mixed-state 1RDM of the steady state is projected to the first excited state rather than the ground state by the purification process (see Section~\ref{section:1RDM}).

\subsection{R\'enyi entropy and entanglement structure of the steady state}

\begin{figure}[t!]
    \centering
    \includegraphics[width=0.5\columnwidth]{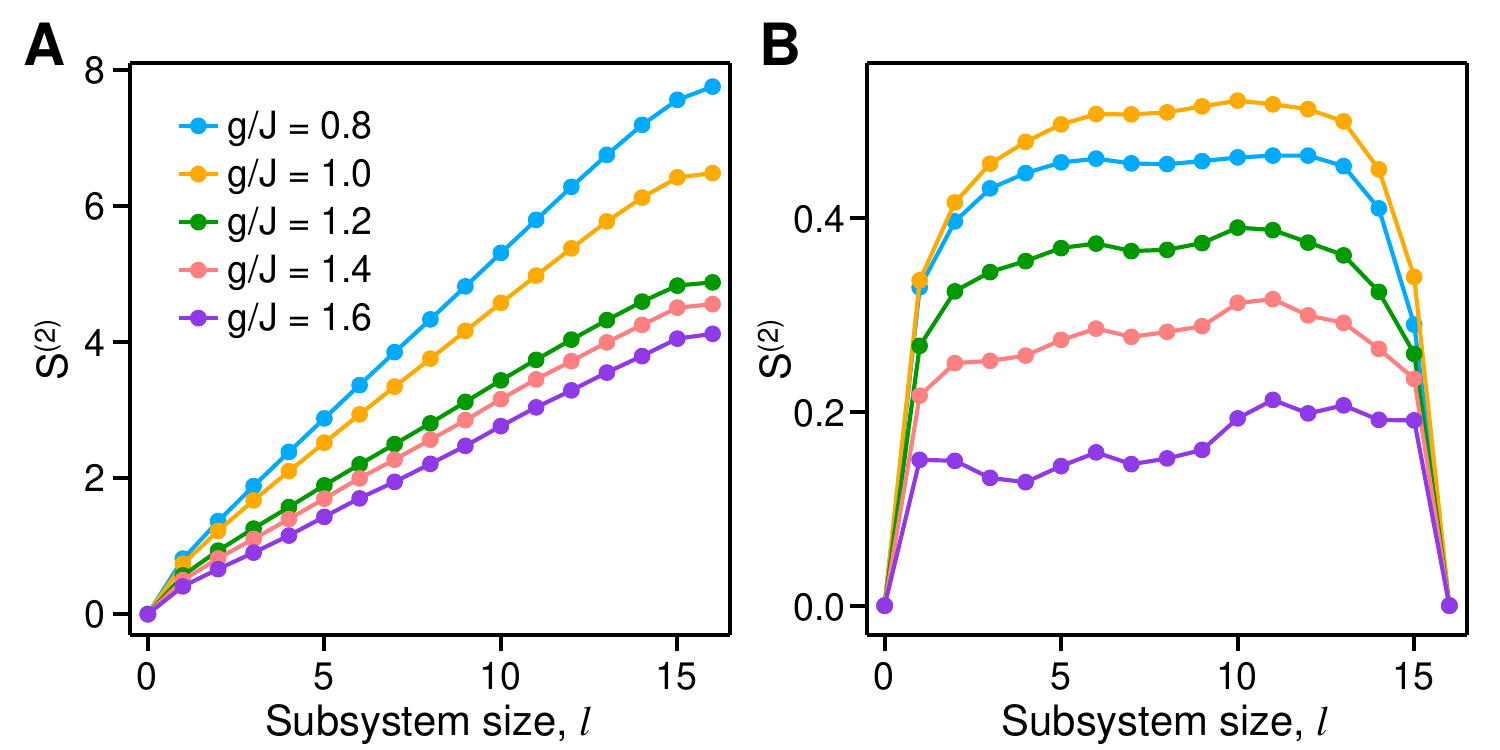}
    \caption{R\'enyi entropy and entanglement structure of the steady state. (A) Second-order R\'enyi entropy $S^{(2)}$ for different subsystem sizes $l$, measured on a 16-qubit chain after it has been dissipatively cooled with $d = 100$ cycles. For each subsystem size, the data are averaged over all possible chains of length $l$. To scramble the steady state, we use 30 sets of randomly chosen single-qubit Clifford gates and perform 3 million measurement shots on each set. (B) Error-mitigated values of $S^{(2)}$ as a function of $l$.}
    \label{fig:s7}
\end{figure}

In addition to quantum correlations, the ground state entanglement structure of the 1D TFIM can also be detected via measurements of the second-order R\'enyi entropy, defined as $S^{(2)} = -\log_2 \text{Tr}{\rho^ 2}$ where $\rho$ is the reduced density matrix of a given subsystem. To measure $S^{(2)}$, we adopt the randomized measurement protocol \cite{Brydges_2019} which has more favorable scaling in the number of measurement shots required compared to full QST. The results for the dissipatively cooled steady state of a $L = 16$ qubit chain is shown in Fig.~\ref{fig:s7}A. Here we observe that $S^{(2)}$ increases nearly monotonically with system size. This is a consequence of an extensive background classical entropy due to the mixed-state nature of the steady state and measurement errors.

Despite the background entropy, past works have shown that it is still possible to extract the entanglement scaling of a quantum state \cite{kaufman_science_2016, hoke2023quantum}. This is because background classical entropy such as measurement error typically scales linearly, with a slope $S^{(2)}_{\text BG}$, against the subsystem size. Since the entanglement entropy is expected to be 0 at the full system size for a pure state, $S^{(2)}_{\text BG}$ = $S^{(2)}_{\text L} / L$ where $S^{(2)}_{\text L}$ is $S^{(2)}$ measured at $l = L$ and has contributions only from the background entropy. The error-mitigated entanglement entropy $S^{(2)}$ for each subsystem is then extracted by subtracting $(l/L)S^{(2)}_{\text BG}$ from the unmitigated value of $S^{(2)}$. The error-mitigated values of $S^{(2)}$ are shown in Fig.~\ref{fig:s7}B. Here we see that the $S^{(2)}$ exhibits area-law scaling while in the paramagnetic phase ($g/J > 1.0$). At the critical point $g/J = 1.0$, $S^{(2)}$ has the largest value and the strongest dependence on system size $l$, consistent with the expected logarithmic scaling of entanglement. In the antiferromagnetic regime ($g/J = 0.8$), $S^{(2)}$ starts to decrease and approach an area law again. These results indicate the entanglement structure of the 1D TFIM is preserved in the dissipatively cooled steady state. An alternative method of extracting entanglement entropy using purified 1RDMs is presented in Fig.~\ref{fig:RDM}, where a similar transition from area-law to logarithmic scaling of entanglement is observed.

\section{Mechanism of dissipative cooling}\label{Sec:Dis}

Here, we discuss the mechanism of dissipative cooling, focusing on the example of the Trotterized, or Floquet TFIM. First, for completeness we provide expressions for the eigenmodes of the Floquet TFIM, obtained by mapping it onto a kicked Kitaev fermionic chain. Second, we introduce an auxiliary at the edge. Assuming a weak coupling between the auxiliary qubit and the chain, we derive a perturbative expression for system's evolution. Adopting secular approximation, we analyze time evolution of quasiparticle occupation numbers. This allows us to identify the parameter values where cooling protocol is optimal and lowest quasiparticle occupations are reached in the steady state. Finally, we illustrate the validity of the secular approximation for a broad range of auxiliary-system couplings, by comparing predicted quasiparticles occupations to exact numerical results. 

\subsection{Eigenmodes of the Floquet transverse-field Ising model}\label{subsection:eigenmodes}

We start by considering the Floquet TFIM described in the main text, which is specified by a cycle unitary operator:
\begin{equation}\label{eq:UTFIM}
\hat{U} = e^{-\frac{i \pi J}{2} \sum_{j=1}^{L - 1} \hat{X}_j \hat{X}_{j + 1}} e^{\frac{i \pi g}{2} \sum_{j=1}^{L} \hat{Z}_j}.
\end{equation}
This model can be mapped onto a quadratic fermionic chain by the Jordan-Wigner transformation. We define Majorana operators on site $j$ as follows: 
\begin{equation}\label{eq:Majoranas}
\hat a_{2j-1}=\left[ \prod_{k=1}^{j-1} \hat Z_k \right]  \hat X_j, \;\;\; \hat a_{2j}=\left[ \prod_{k=1}^{j-1} \hat Z_k \right]  \hat Y_j. 
\end{equation}
These operators obey standard Majorana anti-commutation relations, and are related to complex fermion operators $\hat c_j, \hat c_j^\dagger$ via
$$
\hat a_{2j-1}=\hat c_j^\dagger+\hat c_j, \;\;\; \hat a_{2j}=i\left( \hat c_j^\dagger -\hat c_j \right). 
$$
We note that the fermionic vacuum defined with respect to operators $\hat c_j,\hat c_j^\dagger$, corresponds to $|1\rangle$ state of the qubits/spins. 

The spin operators that enter the expression for the Floquet unitary (\ref{eq:UTFIM}) are related to the Majorana operators as follows, 
\begin{equation}
    \hat Z_j=-i\hat a_{2j-1} \hat a_{2j}, \;\;\; \hat X_j \hat X_{j+1}=-i \hat a_{2j} \hat a_{2j+1}.
\end{equation}
Thus, $\hat{U}$ is a quadratic evolution operator in terms of fermions, and the Majorana operators are linearly transformed under it: 
\begin{equation}\label{eq:eigK}
    \hat U^\dagger \hat a_k \hat{U}=\sum_{l=1}^{2L} K_{kl} \hat{a}_l.
\end{equation}

\begin{figure}
    \centering
    \includegraphics[width=0.5\columnwidth]{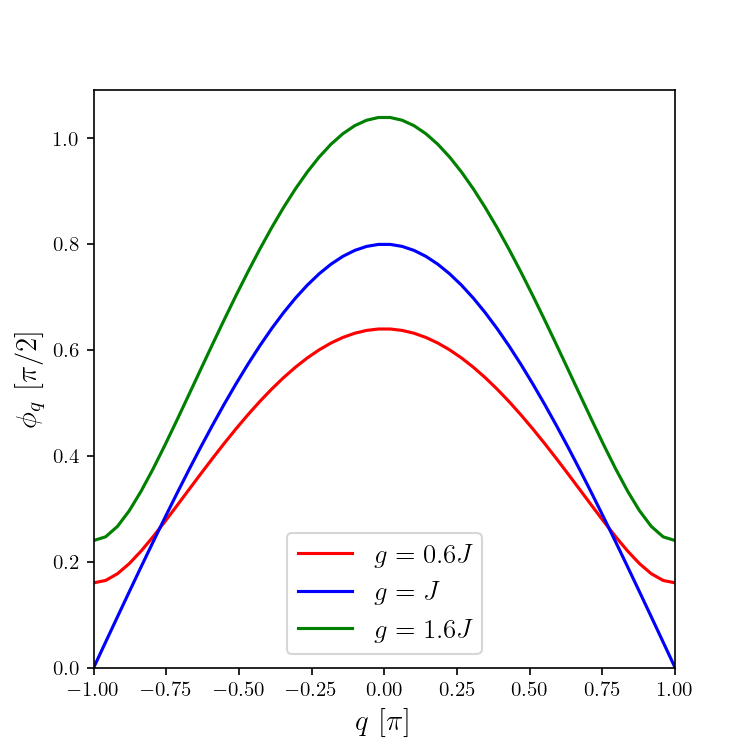}
    \caption{The quasienergy band spectrum $\phi_q$ defined in (\ref{eq:phi_q}) as a function of the quasimomentum $q$, for $J=0.2$, $g$ within antiferromagnetic phase ($g=0.6J$), at critical point ($g=J$) and within paramagnetic phase ($g=1.6J$). At the critical point the band gap closes.}
    \label{fig:quasienergies}
\end{figure}

We look for the eigenmodes of the Floquet TFIM, specified by the annihilation/creation operators $\hat\eta, \hat\eta^\dagger$ such that $\hat U^\dagger \hat\eta \hat U=e^{-i\phi}\hat\eta$ ($\phi$ being the quasienergy), in the following form: 
\begin{equation}\label{eq:eta_expansion}
    \hat\eta=\sum_{j=1}^L \psi_{2j-1} \hat a_{2j-1}+\psi_{2j} \hat a_{2j}. 
\end{equation}
In an infinite system, the solutions have a plane-wave form with quasimomentum $q$, with quasienergy dispersion relation specified by
\begin{equation}\label{eq:phi_q}
  \cos \phi_q= \cos(\pi J) \cos(\pi g)-\sin(\pi J) \sin (\pi g) \cos q. 
\end{equation}
In Fig.~\ref{fig:quasienergies}, we plot the quasienergy bands as a function of $q$ for an infinitely long chain. 

We are interested in the case of a finite chain. In this case, the eigenmodes are given by a superposition of two plane waves with $\pm q$. The boundary conditions yield a transcendental equation for quasimomentum quantization that determines $L$ quasimomenta values, $q_\alpha$. The corresponding quasienergies are specified by Eq.~(\ref{eq:phi_q}). These eigemodes are derived as follows:

For the Floquet TFIM, the action of $K$ on vectors is given by $\underline{v}' = K\underline{v}$, where
\begin{gather}
    v_{2j-1}' = -\sin(\pi J)\sin(\pi g)v_{2j-3}+\sin(\pi J)\cos(\pi g)v_{2j-2}+\cos(\pi J)\cos(\pi g)v_{2j-1}+\cos(\pi J)\sin(\pi g)v_{2j}, \nonumber \\
    v_{2j}' = -\cos(\pi J)\sin(\pi g)v_{2j-1}+\cos(\pi J)\cos(\pi g)v_{2j}-\sin(\pi J)\cos(\pi g)v_{2j+1}-\sin(\pi J)\sin(\pi g)v_{2j+2},\label{eq:bulkeqns}
\end{gather}
for $1< 2j-1 \leq 2L-1$ and the open boundary conditions fix 
\begin{gather}
    v_1' = \cos(\pi g) v_1 +\sin(\pi g) v_2,     \label{eq:boundaryeqn1} \\
    v_{2L}' = -\sin(\pi g) v_{2L-1} + \cos(\pi g) v_{2L} \label{eq:boundaryeqn2}.
\end{gather}
We look for the $L$ eigenvectors of $K$ with non-negative quasienergy $\phi_q \geq 0$ such that $\underline{\psi}'^q = e^{-i\phi_q}\underline{\psi}^q$. Here $q$ labels the quasimomentum. Due to particle-hole symmetry the eigenvectors of $K$ with negative quasienergy are related to those of positive quasienergy by the conjugation $\underline{\varphi}^q = (\underline{\psi}^{q})^*$, $\underline{\varphi}'^q = e^{i\phi_q}\underline{\varphi}^q$. First we derive the plane waves $\underline{v}$ satisfying the eigenvalue equation in the bulk (\ref{eq:bulkeqns}), but not the the boundary conditions (\ref{eq:boundaryeqn1}, \ref{eq:boundaryeqn2}). 

The Bloch ansatz

\begin{equation}
    \begin{pmatrix} v^q_{2j-1} \\ v^q_{2j} \end{pmatrix} = \frac{e^{iq(j-1)}}{\sqrt{L}} \begin{pmatrix} \chi^q_1 \\ \chi^q_2 \end{pmatrix},
\end{equation}
reduces the bulk equations (\ref{eq:bulkeqns}) to the secular equation 
\begin{equation}\label{eq:su2sec}
    \begin{pmatrix}
        \cos(\pi J)\cos(\pi g)-\sin(\pi J)\sin(\pi g)e^{-iq} && \cos(\pi J)\sin(\pi g) +\sin(\pi J)\cos(\pi g)e^{-iq} \\ -\cos(\pi J)\sin(\pi g) -\sin(\pi J)\cos(\pi g)e^{iq} && \cos(\pi J)\cos(\pi g)-\sin(\pi J)\sin(\pi g)e^{iq} \end{pmatrix} \begin{pmatrix} \chi^q_1 \\ \chi^q_2 \end{pmatrix}  = e^{-i\phi_q}\begin{pmatrix} \chi^q_1 \\ \chi^q_2 \end{pmatrix}.
\end{equation}
The two eigenvalues are $e^{-i\phi_q}$ and $e^{+i\phi_q}$ and taking the matrix trace yields Eq. (\ref{eq:phi_q}). The matrix transformation (\ref{eq:su2sec}) can be viewed as an $SU(2)$ rotation by angle $2\phi_q$ about the axis defined by the unit vector
\begin{equation}
    \underline{n}(q) = \begin{pmatrix}
        \sin(2\mu_q)\cos \xi_q \\ \sin(2\mu_q)\sin \xi_q \\ \cos(2\mu_q)
    \end{pmatrix}  \equiv \frac{1}{\sin(\phi_q)} \begin{pmatrix}
        \sin(\pi J)\cos(\pi g) \sin q \\ -\cos(\pi J)\sin(\pi g) - \sin(\pi J) \cos(\pi g) \cos q \\ -\sin(\pi J)\sin(\pi g) \sin q 
    \end{pmatrix},\label{eq:n}
\end{equation}
where we have parameterized by polar and azimuthal angles $\mu_q$ and $\xi_q$, which depend on the quasimomentum. The eigenvector with eigenvalue $e^{-i\phi_q}$ is given by
\begin{equation}
    \begin{pmatrix} \chi^q_1 \\ \chi^q_2 \end{pmatrix} = \begin{pmatrix}
        \cos \mu_q  \\ e^{i\xi_q}\sin\mu_q
    \end{pmatrix}.
\end{equation}
Each quasienergy $\phi_q$ is degenerate with its quasimomentum-reversed partner $\phi_{-q}$ (with the exception of $q=0$ and $q=\pi$, which must be treated separately). The boundary lifts this degeneracy. We then form standing waves $\underline{\psi}^q$ as linear combinations of $\underline{v}^q$ and $\underline{v}^{-q}$, with $\underline{\psi}^q$ satisfying the boundary equations. Introducing the phase shift $\delta_q$, the standing waves take the form
\begin{equation}
    \begin{pmatrix} \psi^q_{2j-1} \\ \psi^q_{2j} \end{pmatrix} = \frac{1}{\sqrt{L}}\begin{pmatrix} e^{i\delta_q}\chi^{-q}_1 && \chi^q_1 \\ e^{i\delta_q}\chi^{-q}_2 && \chi^q_2 \end{pmatrix}  \begin{pmatrix} e^{-iq(j-1)} \\ e^{iq(j-1)} \end{pmatrix},
\end{equation}
where $\delta_q$ is chosen so as to satisfy the left boundary condition (\ref{eq:boundaryeqn1}),
\begin{equation}
    e^{i\delta_q} = \frac{-(e^{-i\phi_q}-\cos(\pi J))+e^{i\xi_q}\sin(\pi g)\tan\mu_q}{(e^{-i\phi_q}-\cos(\pi J))\tan\mu_q+\sin(\pi g)e^{-i\xi_q}}. \label{eq:delta-q}
\end{equation}
The right boundary condition (\ref{eq:boundaryeqn2}) yields a transcendental equation for the quasimomenta quantization:
\begin{equation}\label{eq:quantization}
    e^{2iq(L-1)} = - \bigg[\frac{(e^{-i\phi_q}-\cos(\pi J))-e^{i\xi_q}\tan \mu_q\sin(\pi g) }{(e^{i\phi_q}-\cos(\pi J))e^{i\xi_q}\tan \mu_q +\sin(\pi g)}\bigg]^2.
\end{equation}
In the limit of large $L$ and small quasimomentum $q \ll \pi$, Eq. (\ref{eq:quantization}) can be replaced by the usual formula
\begin{equation}
    q_\alpha \sim \frac{\pi(\alpha-1)}{L}, \hspace{.5cm} \alpha \in (1, L).
\end{equation}
The standing wave solutions $\underline{\psi}^q$ are the operator coefficients appearing in Eq. (\ref{eq:eta_expansion}), for the eigenmode $\hat\eta^q$.

\subsection{Perturbation theory for Floquet evolution}

Here, we provide a perturbative expression for the state of a Floquet system coupled to an auxiliary, assuming that the auxiliary-system coupling is weak. We start by considering a general setup, where the system and auxiliary first undergo $M$ periods of unitary evolution, specified by an operator
\begin{equation}
    \hat{\mathcal{U}}=\hat U_{\rm SA} \hat U_{\rm A} \hat U_{\rm S}, 
\end{equation}
followed by the reset of auxiliary to a state $\rho_{\rm A}^0$. The auxiliary-system coupling is chosen to be in the form
\begin{equation}
    \hat U_{\rm SA}=e^{i\theta \hat K},
\end{equation}
with $\hat K$ being system-auxiliary coupling Hamiltonian. We will be interested in the limit of weak coupling, $\theta\ll 1$. 

To find the density matrix of the system after one dissipative cycle (${\mathcal{U}}^M$ followed by auxiliary reset), it is convenient to use interaction representation for operators: 
\begin{equation}
    \hat A_{I}(s)=\hat U_0^{-s} \hat A\hat U_0^{s}, 
\end{equation}
where $s$ is the discrete time (number of unitary evolution periods within one dissipative cycle, such that $s\in [0;M]$) and $\hat U_0=\hat U_{\rm A}\hat U_{\rm S}$ is the unperturbed evolution operator. The unitary evolution operator can be written as
\begin{equation}\label{eq:evolutionM}
    {\hat{\mathcal{U}}}^M=\hat U_0  \mathcal{T} \prod_{s=1}^M e^{i\theta \hat K_I (s)}, 
\end{equation}
where $\mathcal{T}$ denotes time-ordering. 

Next, we focus on the system's density matrix after one dissipative cycle. At the beginning of the cycle, the system and auxiliary are described by a density matrix $\rho^{(n)}\otimes \rho_A^{0}$, where $\rho_A^0$ is the state to which the auxiliary is reset.  Expanding the evolution operator in Eq.(\ref{eq:evolutionM}) to second order in $\theta$, and tracing out the auxiliary, we obtain system density matrix in the interaction representation: 
\begin{equation}\label{eq:DMevolution}
   \hat U_{\rm S}^{-M} \rho^{(n+1)} \hat U_{\rm S}^M =  \rho^{(n)} -\theta^2 \sum_{\substack{s_2=1 \\ s_1 < s_2}}^{M} {\rm Tr}_A [ \hat K_I (s_2), [\hat K_I (s_1), \rho^{(n)}\otimes \rho_{\rm A}^0]  ]-\frac{\theta^2}2 \sum_{s=1}^M {\rm Tr}_A [\hat K_I (s), [\hat K_I(s), \rho^{(n)}\otimes \rho_{\rm A}^0]]. 
\end{equation}
Next, we define the density matrix in the interaction representation with respect to system only evolution, to describe system's state after many dissipative cycles: 
\begin{equation}\label{eq:SystemIntRepr}
    \rho^{(n)}_{\rm int}\equiv \hat U_{\rm S}^{-Mn} \rho^{(n)} \hat U_{\rm S}^{Mn}.
\end{equation}
The advantage of considering the density matrix in the interaction representation is that the change of $\rho_{\rm int}$ over one dissipative cycle is proportional to $\theta^2$, and therefore small provided $\theta\ll 1$. This change is obtained from Eqs.(\ref{eq:DMevolution},\ref{eq:SystemIntRepr}). 

\subsection{Application to the Floquet TFIM: secular approximation}

Next, we analyze the cooling of the Floquet TFIM, described in the main text. In this case, 
\begin{equation}
    \hat U_{\rm A}=e^{i\frac{\pi h}2 \hat Z_a}, \;\;\; \hat K=\frac{1}2 \left(\hat X_a \hat X_1 + \hat Y_a \hat Y_1\right). 
\end{equation}
Since our purpose here is mostly to illustrate the cooling mechanism, for simplicity we consider an auxiliary coupled to the first site of the chain.

We perform Jordan-Wigner transformation, described above, arriving at the following Floquet operator, written in terms of fermionic eigenmodes of the chain $\hat\eta_k $ with quasienergies $\phi_k$ (see above), and in terms of fermionic operator $\hat d$ acting on the auxiliary site: 
\begin{equation}\label{eq:USA_Ising}
    \hat U_{\rm S}=e^{-i\sum_k \phi_k \hat\eta_k^\dagger \hat\eta_k}, \;\;\; \hat U_{\rm A}=e^{-i{\pi h}\hat d^\dagger \hat d}, \;\;\; \hat K=\hat d^\dagger \hat c_1 +\hat c_1^\dagger \hat d, 
\end{equation}
where $\hat c_1$ is the annihilation operator on the first site of the chain, introduced above. We further express the operator $\hat c_1$ via eigenmode operators $\hat\eta_k, \hat\eta_k^\dagger$: 
\begin{equation}\label{eq:c1eigenmodes}
    \hat c_1=\sum_k \alpha_k \hat\eta_k +\beta_k \hat\eta_k^\dagger, \;\;\; \hat c_1^\dagger=\sum_k \alpha_k^* \hat\eta_k^\dagger +\beta_k^* \hat\eta_k . 
\end{equation}
The coefficients $\alpha_k,\beta_k$ are obtained from the expressions for the eigenmode wave functions. 

In the experiment, we reset the auxiliary to $|0\rangle$ state, which corresponds to the occupied $d$-level in the fermionic language. Thus, 
\begin{equation} \label{eq:rhoA0}
{\rm Tr}_A( \hat d^\dagger \hat d \rho_{\rm A}(0))=1, \;\;\; {\rm Tr}_A( \hat d \hat d^\dagger  \rho_{\rm A}(0))=0. 
\end{equation}
From Eqs.(\ref{eq:c1eigenmodes},\ref{eq:USA_Ising}), we obtain the expression for the operator $\hat K$ in the interaction picture, 
\begin{equation}
    \hat K_I (s)=\sum_k  \alpha_k e^{i(\pi h-\phi_k) s} \hat d^\dagger \hat\eta_k +\beta_k e^{i(\pi h+\phi_k) s} \hat d^\dagger \hat\eta_k^\dagger + {\rm h.c.}, 
\end{equation}
 where ${\rm h.c.}$ denotes hermitian conjugate. Combining this equation with Eq.(\ref{eq:DMevolution}) and Eq.(\ref{eq:rhoA0}), we obtain system's density matrix evolution. 

 Further, we consider the density matrix in the interaction representation (\ref{eq:SystemIntRepr}). This leads to dressing of the fermionic operators $\hat\eta_k^\dagger, \hat\eta_k$ entering the equation for the change of density matrix, $\Delta \rho_{\rm int}=\rho_{\rm int}^{(n+1)}-\rho_{\rm int}^{(n)}$, by phases $e^{\pm iMn\phi_k}$, respectively. 

 Next, we adopt the standard secular approximation: assuming $\theta^2 \ll \delta \phi$, where $\delta\phi$ is the quasienergy level spacing, we coarse-grain the time evolution of $\rho_{\rm int}^{(n)}$ over a number of dissipative cycles of order $(M\delta \phi )^{-1}$, and observe that the terms of the form $\hat\eta_k^\dagger \hat\eta_q$ with $k\neq q$ can be neglected due to their oscillating phases. Thus, we are left only with the contributions where $k=q$. This results in a simplified equation for the density matrix evolution (over one dissipative cycle):
\begin{align}
\frac{d\rho _{\text{int}}^{(n)}}{dn}-i \left[\rho_{\text{int}}^{(n)},\Delta H_{\text{S}}\right]=\nonumber
\end{align}
\vspace{-0.2 in}
\begin{align}
+\sum _{k = 1}^L W^+\left(q_k\right)\left(\eta _k^{\dagger }\rho_{\text{int}} \eta _k-\frac{1}{2}\left\{\eta _k\eta _k^{\dagger },\rho_{\text{int}}\right\} \right)+W^-\left(q_k\right)\left(\eta _k\rho_{\rm int} \eta _k^{\dagger }-\frac{1}{2}\left\{\eta
_k^{\dagger }\eta _k,\rho_{\text{int}}\right\}\right)\;,\label{eq:DM_secular}
\end{align}
where $\{A,B\}$ denotes anticommutator of the operators $A$ and $B$ , the sum in the r.h.s. is over  $L$ fermionic  eigenmodes. $W^{+}(q)$ and  $W^{-}(q)$ are probabilities to, respectively,  create and annihilate a fermion mode $k$ with an absolute value of the quasimomentum $q_k$ over one  dissipative cycle. We express these quantities via the properties of the eigenmodes of the Floquet TFIM described above, by relating the coefficients $\alpha_k, \beta_k$ in Eq.~(\ref{eq:c1eigenmodes}) to the eigenmode amplitudes $\psi_{2j-1}^q, \psi_{2j}^q$:
  \begin{equation}
W^{\pm }(q)\equiv 2\pi  M \theta ^2\,\left| \psi _1^q\pm i \psi _2^q \right| {}^2 \delta _M\left( \phi_q \pm  \pi h\right),\label{eq:Wpm}
\end{equation} \begin{equation}
\left| \psi _1^q\pm i \psi _2^q \right|^2=\frac{1}{L}f_q^\pm,\qquad  f_q^\pm=4\left| \cos \left(\mu _q\right) \cos \left(\frac{\delta _q}{2}\right)\mp  \sin \left(\mu
_q\right)\sin \left(\xi _q-\frac{\delta _q}{2}\right)\right|^2\;.\label{eq:fq}
\end{equation}
where $\mu_q$ , $\xi_q$ are defined via Eq.~(\ref{eq:n}) and the phase shifts  $\delta_q$ are is given in Eq.~(\ref{eq:delta-q}).

The equation (\ref{eq:DM_secular}) takes the form of the Lindblad equation quadratic in fermion operators. It  has a simple physical meaning: the first term in the r.-h.s. describes quasiparticles being removed from the system, with a rate $W^-(q)$ that depends on the weight of the quasiparticle wave function with momentum $q$ on the site coupled to the auxiliary, and on the phase difference $\pi h+\phi_q$. Similarly, the second term describes processes where quasiparticles are being excited from the vacuum. 

 In Eq.~(\ref{eq:DM_secular})  $\Delta H_{\text{S}}$ is effective Hamiltonian correction  (collective "Lamb shift") of  the system produced by the auxiliary qubits
 \
  \begin{equation*}
\Delta H_{\text{S}}=\frac{M \theta ^2}{L} \sum _{k = 1}^L \Delta \left(q_k\right)\eta _k^{\dagger }\eta _k-\Delta _0
\end{equation*}
\begin{equation*}
\Delta (q)=\sum _{\sigma =\pm } f_q^{\sigma }\mathcal{P}_M\left(\frac{1}{\phi _q+ \sigma \pi h}\right) ,\hspace{8 mm}\Delta _0=\frac{M \theta
^2}{L}\hspace{2 mm}\sum _{\mu  = 1}^L \hspace{2 mm}f_q^+\mathcal{P}_M\left(\frac{1}{\mathcal{E}_{\mu }+ \pi h}\right)
\end{equation*}

 The functions $\delta_M(x)$ and $\mathcal{P}_M(x)$ above  approximate delta function and principle value function in the limit $M\rightarrow \infty$
  \begin{equation*}
\delta _M(x)=\frac{1}{2\pi M}\frac{ \sin \left(\frac{M x}{2}\right)^2}{\sin \left(\frac{x}{2}\right)^2},\hspace{8 mm}\mathcal{P}_M\left(\frac{1}{x}\right)=\frac{1}{M}\sum
_{m=1}^M \sum _{l=1}^{m-1} \sin (xl)
\end{equation*}
 (for brevity we omit  the explicit form of the double sum).
 From the fermionic Lindblad equation (\ref{eq:DM_secular})  one can n obtain the steady-state population of the quasiparticle levels  given by: 
 \begin{equation}\label{eq:QP_occupation}
n_k=\left\langle \eta _k^{\dagger }\eta _k\right\rangle =\left(1+\frac{f_q^-}{f_q^+}\frac{\delta _M\left(\pi h-\phi _q\right)}{\delta _M\left(\pi
h+\phi _q\right)}\right)^{-1}
\end{equation}
This analytical expression allows one to theoretically identify the optimal value of the parameter $h$, for which the steady-state quasiparticle number is minimal. As a test, we calculate an optimal value of $h = 1.60$ for the model parameters in Fig.~\ref{fig:s4}a, close to the experimentally determined value of $h = 1.65$. This coincides with the upper edge of the quasiparticle band.

\begin{figure*}[t!]
    \centering
    \includegraphics[width=\columnwidth]{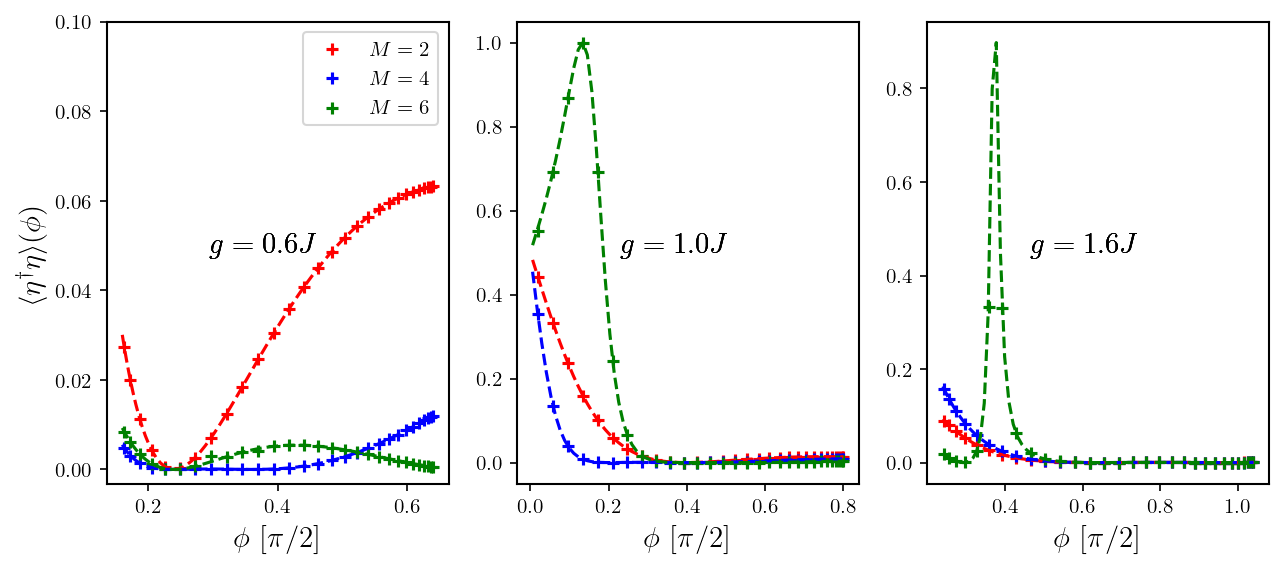} 
    \caption{Quasiparticle occupations in the steady state, as a function of the quasiparticle quasienergy $\phi$: a comparison of secular approximation prediction (dashed lines) and numerical results (crosses). The coupling of auxiliary and the system is chosen to be weak, $\theta/\pi=0.001$, and $M$ denotes a number of cycles before auxiliary is reset. The ancilla field $h$ is tuned to the approximately optimal upper band edge, $h=J+g$. The system size is $L=30$ sites. Parameter $J=0.2$. We observe a quantitative agreement between the two approaches, with the lowest quasiparticle population achieved for $M=4$. We note that the quasiparticle population at zero quasienergy remains large at the critical point (middle panel).}
    \label{FigSecular}
\end{figure*}

\begin{figure*}[t!]
    \centering
    \includegraphics[width=\columnwidth]{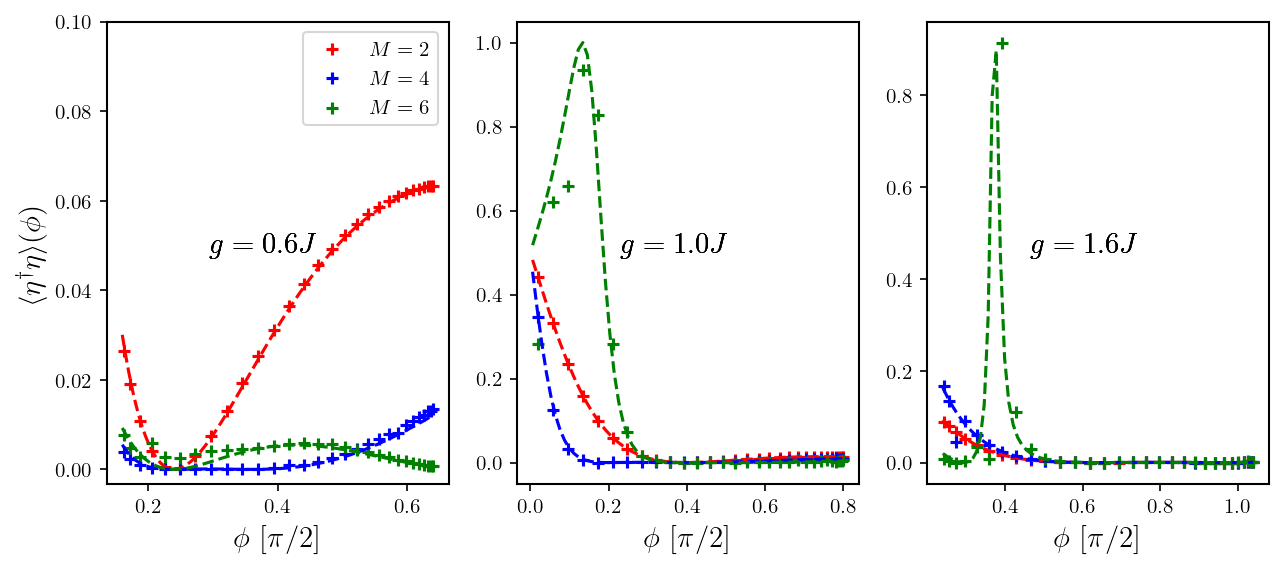} 
    \caption{Same as in Fig.~\ref{FigSecular}, but with stronger auxiliary-system coupling  $\theta/\pi=0.01$. Despite visible deviations from an exact result, the secular approximation qualitatively captures the behavior of quasiparticle population in the steady state.}
    \label{FigSecularLarge}
\end{figure*}

As a next step, it is instructive to verify the validity of the secular approximation. To this end, we first compare the analytical prediction (\ref{eq:QP_occupation}) with the exact numerical calculation. The result for weak auxiliary-system coupling, $\theta/\pi=0.001$, is illustrated in Fig.~\ref{FigSecular}. We use the value $h=J+g$ for ancilla field, which agrees with the optimal value above. We observe excellent agreement in all regions of the phase diagram and for different values of unitary evolution periods $M$ before a reset. 

Further, we study the case of stronger coupling $\theta/\pi=0.01$ (Fig.~\ref{FigSecularLarge}). Interestingly, at the critical point $g=J$, the secular approximation remains accurate. In the two gapped phases, the approximation captures qualitative features, but significant deviations from exact results are visible, especially near the band edges. Nevertheless, secular approximation provides a good guide for identifying optimal auxiliary parameters.

\section{1RDM of the Ising chain and purification}\label{section:1RDM}

\begin{figure}[t!]
\includegraphics[width=1\columnwidth]{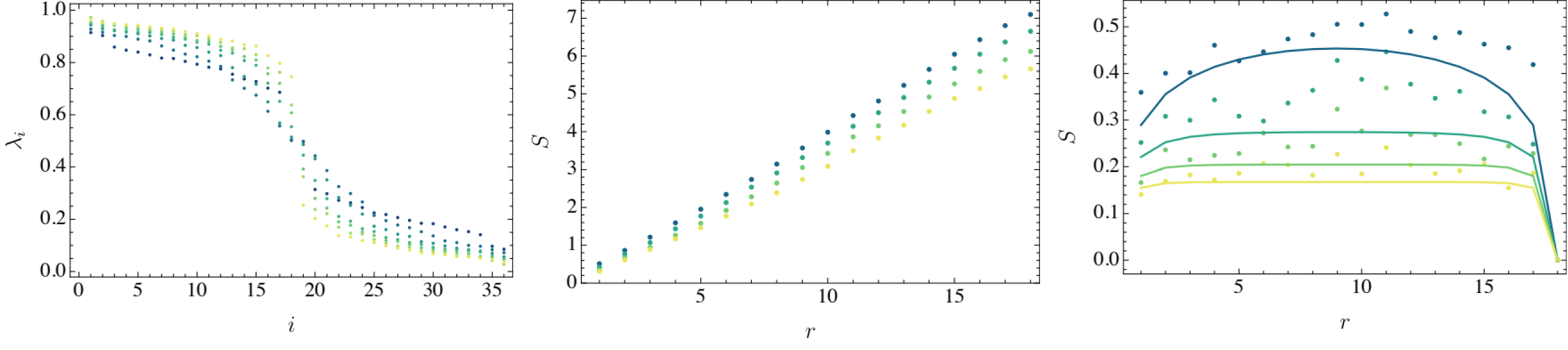}
\caption{Left panel: Eigevalues of the 1RDM, $D$, in the NESS of the Kicked Ising model for $L= 18$ qubits. The parameters are $(g, J) = (0.08, 0.2)$ and  $g=(0.6,0.8,\ldots 1.6) J$ for $J = 0.25 $. Lighter colours denote increasing $g$. Middle panel: The dependence of entanglement entropy for a quadratic fermionic system described by the experimental 1RDM, $D$. We only plot the parameters $g/J\geq 1$. Right panel: Same as before for the purified 1RDM, $D_{pure}$. Full lines correspond to the values for the exact vacuum of the Kicked Ising model, Eq.~(\ref{eq:UTFIM}).
 }
\label{fig:RDM}
\end{figure} 

In this section we describe the one-body density matrix (1RDM) formalism  and the corresponding purification scheme.  Due to the quadratic nature of the Floquet transverse-field Ising model, Eq.~(\ref{eq:UTFIM}), all the information about its many-body eigenstates is contained in the two-body fermionic correlation functions. The latter require a polynomial number of measurements in the system size ($\propto L^2$). The $2 L \times 2 L$ matrix of such correlation functions is referred to as 1RDM. It is conveniently expressed via Majorana operators defined in Eq.~(\ref{eq:Majoranas}):
\begin{equation}\label{Eq:1RDM}
 D = \frac{1}{2}\left(\begin{array}{cc}
D^{oo} & D^{oe}\\
  D^{eo}& D^{ee},
\end{array}\right),\quad D^{oo}_{i,j} = \langle a_{2i-1}a_{2j-1}\rangle, \quad  D^{oe}_{i,j} = \langle a_{2i-1}a_{2 j}\rangle,\quad  D^{eo}_{i,j} = \langle a_{2 i}a_{2j-1}\rangle,\quad D^{ee}_{i,j} = \langle a_{2i}a_{2j}\rangle,\quad i,j \in \{1,L\}.
\end{equation}
Here the averaging is taken over the system's state, described by a density matrix $\rho$: $\langle \cdot \rangle \equiv \text{tr}(\cdot \rho)$. For quadratic states there is a one-to-one relation between the many-body density matrix of the system and 1RDM~\cite{PeschelJPA03}. For many-body states the 1RDM is just a correlation matrix, however we keep the terminology unchanged for clarity.

 The experimentally extracted 1RDM can be written in the basis of eigenmode operators, related to the Majorana operators via Eq.~(\ref{eq:eta_expansion}):
\begin{equation}\label{Eq:1RDMGdG}
    F_{i j}=\left(\begin{array}{cc}
F^{+-} & F^{--}\\
  F^{++}& F^{-+},
\end{array}\right),\quad F^{+-}_{i,j} = \langle \eta^\dag_i \eta_j\rangle, \quad  F^{++}_{i,j} = \langle \eta^\dag_i \eta^{\dag}_j\rangle,\quad F^{-+}_{i,j} = \langle \eta_i \eta^{\dag}_j\rangle,\quad F^{--}_{i,j} = \langle \eta_i \eta_j\rangle,
\end{equation}
 In particular, the quasiparticle occupations are given by $F^{+-}_{i,i} = \langle \eta^{\dag}_i \eta_i\rangle$. 

To purify a 1RDM, we approximate it by the 1RDM, $D_{pure}$, of the closest pure quadratic fermionic state, i.e. a Slater determinant wavefunction. Taking into account the fact that the 1RDM of a Slater determinant wavefunction is a projector, we can express the purification as a constrained minimization problem,
\begin{equation}
    \text{min} |D - D_{pure}|_F, \;\;\;\; {\text{tr}{D_{pure}} = L,D^2_{pure} =D_{pure}},
\end{equation}
 where$| \cdot |_F$ denotes the Frobenius norm. The minimization constraints for the matrix are fixed trace and being a projector. In our approach, we use a purification scheme which is equivalent to the purification proposed by McWeeny~\cite{McWeeny_1960}. The purified 1RDM has the form $D_{pure} = \sum^{L}_{i=1}|i\rangle \langle i|$, corresponding to the projector to the space spanned by the eigenvectors associated to the $L$ largest eigenvalues of the original 1RDM, 
 \begin{equation}
     \text{spec}(D) = \lambda_i, \quad i\in\{1,2L\}, \quad \lambda_i\geq \lambda_{i+1}. 
 \end{equation}
The purified state can be thought of as a  state with occupied fermionic modes $\tilde\eta_i, \tilde\eta_i^\dagger$, which correspond to the $L$ eigenvectors of 1RDM with the largest eigenvalues. The many-body fidelity of the purified state with respect to the ground state, shown in Fig.3C of the main text, is then given by an overlap of two Slater determinant states, defined by sets of modes $\{\tilde\eta_i \}_{i=1}^L$ and $\{\eta_i \}_{i=1}^L$, respectively. This procedure is used to obtain the fidelity illustrated in Fig.3C of the main text.

 In Fig.~\ref{fig:RDM}, we illustrate the effect of purification on the experimentally measured steady state 1RDM for the Floquet TFIM defined in Eq.~(\ref{eq:UTFIM}). We first discuss the properties of the experimentally measured 1RDM, $D$, including eigenvalue spectrum (left panel in Fig.~\ref{fig:RDM}) and entanglement entropy, computed for a quadratic state that corresponds to $D$ (middle panel of Fig.~\ref{fig:RDM}). In the paramagnetic phase $g/J > 1$, a clear ``jump'' in the eigenvalue magnitude at $i = L$ indicates the proximity to a pure state. In the anti-ferromagnetic phase $g/J < 1$, we observe the close degeneracy of the eigenvalues $\lambda_{L}\sim\lambda_{L-1}$.  This reflects the presence of a degenerate ground state manifold due to the presence of a Majorana edge mode. The cooling algorithm leads to a steady state in which the steady state contains a mixture of the two nearly degenerate ground states; this corresponds to the Majorana edge mode being occupied with probability close to $1/2$. To exclude the effect of the Majorana edge modes, we divide the full fidelity by the contribution of the $L$th mode: $\text{Fidelity}\rightarrow \text{Fidelity}/\langle \eta_L  \eta^{\dag}_L \rangle$. This leads to an improved fidelity in the antiferromagnetic phase, illustrated by dashed lines in Fig.3C of the main text.

We note that the high fidelity of the purified state indicates that the modes $\tilde\eta_i$ are close to the true quasiparticle modes $\eta_i$ of the system. In other words, 1RDM in Bogoliubov-de-Gennes basis, Eq.~(\ref{Eq:1RDMGdG}), is almost diagonal. The near-diagonal nature of the density matrix in the quasiparticle basis is justified in the limit of the weak system-ancilla coupling $\theta$, (see discussion about the validity of the secular approximation, Eq.~(\ref{eq:DM_secular}), in the previous section).

 Next we focus on entanglement entropy $S_A = -\text{tr}\rho_B \log \rho_B$, where $\rho_B= \text{tr}_A \rho$ is the reduced density matrix for a partition of the system $A\bigcup B =\{1,2,\ldots,r\}\bigcup \{r+1,\ldots,L\}$. Experimentally determining the full many-body density matrix of the system is prohibitively expensive for large systems, as it scales exponentially with the number of qubits. On the other hand, the 1RDM can be efficiently extracted as it just involves two-point correlation functions. For this reason, by only using experimental data, we calculate  the entanglement entropy of a quadratic system with the same 1RDM as the experimental steady state~\cite{PeschelJPA03}. Even though the exact entanglement entropy of the NESS can be significantly different from our calculation, the 1RDM entanglement entropy nicely illustrates the effect of purification: The volume-law entanglement scaling, $S\propto r$ of the original 1RDM, arising due to decoherence, changes to an area-law scaling, $S \propto \text{const.}$, for the purified 1RDM, for all parameters except at the critical point, $g\neq J$. Additionally we see that the value of the entropy is close to that of the exact vacuum of the TFIM, even at the critical point. This means that critical properties such as the long-range order shown in the main text are also captured by $D_{pure}$.

\section{Comparison between dissipative and unitary state preparation protocols}

In this section we compare the dissipative cooling protocol to the unitary preparation protocol for fermionic Gaussian states proposed by Jiang et al.~\cite{JiangPRA18}. In the absence of decoherence, the unitary protocol efficiently prepares the exact vacuum state of a given quadratic fermionic system. However, weak decoherence present in NISQ devices leads to errors in the prepared state. We explore how the states prepared using the unitary protocol $\rho_U$ compare to the dissipatively cooled states, $\rho_D$.

\begin{figure}[h!]
    \centering
    \includegraphics[width=1\columnwidth]{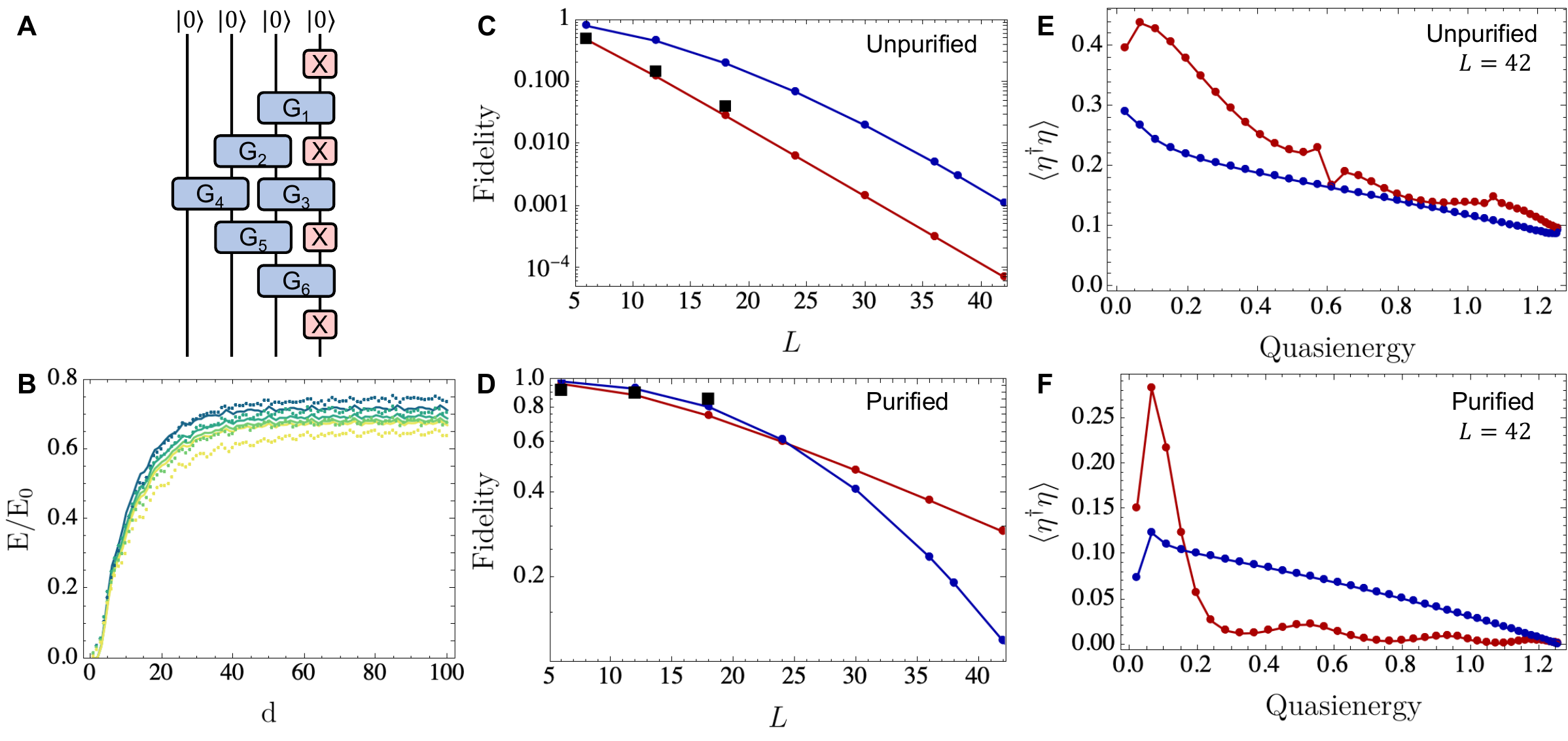}
    \caption{(A) Illustration of the state preparation protocol~\cite{JiangPRA18}, for a system of $L = 4$ qubits. Every qubit is initialized in its ground state. (B) Comparison of energy convergence between experimental data (points) and simulations (lines) at the critical point ($g = J = 0.2$) of the Floquet TFIM (Eq.~(\ref{eq:UTFIM})). System sizes $ L = \{12, 18,24,30\}$ are represented by blue to yellow colors. (C) Fidelity between the exact vacuum state of Floquet TFIM unitary $\hat{U}$ at the aforementioned critical point, and the  states prepared by simulating the dissipative (red) and unitary (blue) protocols. Black squares denote the experimental values for $\{6,12,18\}$ qubits. (D) Same as (C), for the purified states  according to the method described in Section~\ref{section:1RDM}. (E,F) Unpurified and purified quasiparticle occupations for the simulated protocols. We observe that the  purified states generated by the dissipative protocol have considerably lower high-energy quasiparticle occupations.}
    \label{fig:cooling}
\end{figure}

The Gaussian state preparation protocol for a system of $L$ fermions consists of $O(L)$ layers of one- and two-body gates as illustrated in Fig.~\ref{fig:cooling}A. The gates have the form,
\begin{equation}
    G_{n} (\theta_n ,\phi_n) = e^{i \frac{\phi_n}{2} \hat Z_{i_n}}e^{-i \frac{\theta_n}{2}\left(\hat X_{i_n} \hat Y_{i_{n+1}} -\hat Y_{i_n}\hat X_{i_{n+1}}\right)},\qquad B = \hat X_L,
\end{equation}
where the gate index $n \in [1, L (L-1)/2]$, and $i_n$ denotes the position of the corresponding qubit in the system. The angles $(\theta_n,\phi_n)$ are determined from the TFIM unitary $\hat{U}$ (Eq.~(\ref{eq:UTFIM})) by employing the algorithm proposed in~\cite{JiangPRA18}. The weak decoherence in the system is modeled according to Eq.~(\ref{eq:decoherence}). Since the decoherence strength is proportional to the experimental time required to apply the quantum circuit, we apply $\mathcal{D}(\gamma_\theta,\gamma_\phi)$ after every layer of unitary gates. The decoherence rates are set to the qubit coherence rates $\gamma_\theta = 1/T_2\sim 0.016$, $\gamma_\phi = 1/T_1\sim 0.006$, which were previously shown (Fig.~\ref{fig:s4}.B) to match the experimental data.

In order to explore large system sizes we perform the state preparation protocols using tensor-networks techniques. The state is represented by a matrix product density operator (MPDO)~\cite{VerstraetePRL04}. The time integration is performed  by a time-evolving block 
decimation (TEBD) algorithm~\cite{VidalPRL03}, implemented using ITensor library~\cite{itensor}. The bond dimension is set to $\chi = 300$, as this value is found to give converged numerical results for all cases studied.

We focus on the critical point $J = g = 0.2$ of the TFIM, since the long-range order present in the critical ground state is expected to be most susceptible to noise effects, challenging the performance of the preparation protocols. In Fig.~\ref{fig:cooling}B we compare the energy convergence of the numerically simulated dissipative cooling protocol and the experimental data (Fig.~\ref{fig:s5}).  We observe a close agreement for system sizes $L = 12,18,24$ and a slightly worse agreement for $L = 30$. The agreement of these results provides a justification for the choice of the decoherence strength in the protocol comparison.

In Fig.~\ref{fig:cooling}C we show the fidelity between the prepared state and the vacuum state. The unitary preparation yields higher fidelity for the available system sizes, $\langle \rho_D \rangle < \langle \rho_U \rangle$. However, we expect that for larger system sizes, where the number of layers required for the unitary protocol requires running times that are much longer than the qubit coherence times, the dissipative protocol will become more efficient. Next, in Fig.~\ref{fig:cooling}D we present the fidelities following purification (see Section~\ref{section:1RDM}) of the states, $\rho^P_D,\rho^P_U$. We observe that $\langle\rho^P_U \rangle$ decays considerably faster than $\langle\rho^P_D \rangle$ and for $L \geq 25$ onward, $\langle\rho^P_D \rangle > \langle\rho^P_U \rangle$, illustrating a better performance of the dissipative cooling algorithm. 

To further understand the structure of the approximate vacuum states prepared by the two protocols, we calculate the density of quasiparticle excitations in the system, Fig.~\ref{fig:cooling}E,F. The density of quasiparticles in $\rho_D$ depends strongly on the quasiparticle quasienergy, while for $\rho_U$ this is not the case. In addition, we observe that excitations at sufficiently high quasienergies are suppressed by purification more efficiently for $\rho_D$. This is a result of the different 1RDM structure in the quasiparticle basis, Eq.~(\ref{Eq:1RDMGdG}), for the two protocols. The dissipatively prepared state $\rho_D$ is close to a diagonal mixture of different quasiparticle states. In contrast, the density matrix $\rho_U$ reached by the unitary protocol features larger off-diagonal matrix elements. For this reason, the purification scheme performs considerably better on $\rho_D$.

\section{Transport in Floquet XXZ under maximal pumping}

In this section of the Supplementary Material (SM), we provide numerical simulations of the non-equilibrium quantum transport in the Floquet XXZ chain and compare them to the experimental results. 

\subsection{Model and setup}

For the driving protocol we use two auxiliary qubits coupled to the boundaries of the system of $L$ qubits. We label the qubits according to the definitions of Fig. 1 of the main text: The left and right auxiliaries are denoted by $Q_{a,1}$ and $Q_{a,2}$, respectively. The qubits of the system are denoted as $Q_{s,1}, \ldots , Q_{s,N-2}$. The system size is therefore $N = L + 2$.

Our system is inspired by the XXZ-Hamiltonian,
\begin{equation}
H_{XXZ} = \sum^{L-1}_{i=1}h_i,\quad  h_i =\theta \left(\sigma^+_i\sigma^-_{i+1} + \text{h.c.}\right) +\phi n_i n_{i+1},
\end{equation}
where $i$ denotes the position of the system qubit in the chain, $n=|1\rangle\langle 1|$ is the particle density and $\sigma^\pm$ are the hardcore boson creation/annihilation operators. Similarly to the Floquet system (see main text), the anisotropy parameter $\Delta = \frac{\phi}{2\theta}$ controls the transport properties of the Hamiltonian system~\cite{ProPRL11,BertiniRMP2021}. 

The Floquet XXZ chain is realized by a trotterization of the Hamiltonian evolution,
\begin{equation}\label{eq:XXZ}
    U_{XXZ} =  U_{even}U_{odd},\hspace{10mm}    U_{even}= \prod^{N/2}_{i=1}\text{FSim}_{2i,2i+1}, \hspace{10 mm} U_{odd}= \prod^{N/2-2}_{i=1}\text{FSim}_{2i+1,2i+2}, \hspace{10 mm} .
\end{equation}
where the two-qubit gates are generated by the Hamiltonian density as
\begin{equation}
   \text{FSim}_{i,i+1}(\theta ,\phi) = \exp\left(
   i h_i(\theta,\phi) \right). 
   \end{equation}
The Floquet XXZ chain retains the integrable structure of the Hamiltonian model, and therefore, features a macroscopic number of conserved quantities~\cite{Gritsev17integrable}. In addition, the total number of particles $N_{tot} = \sum_i n_i$ in the system is conserved. 

\begin{figure}[t]
\includegraphics[width=0.8\columnwidth]{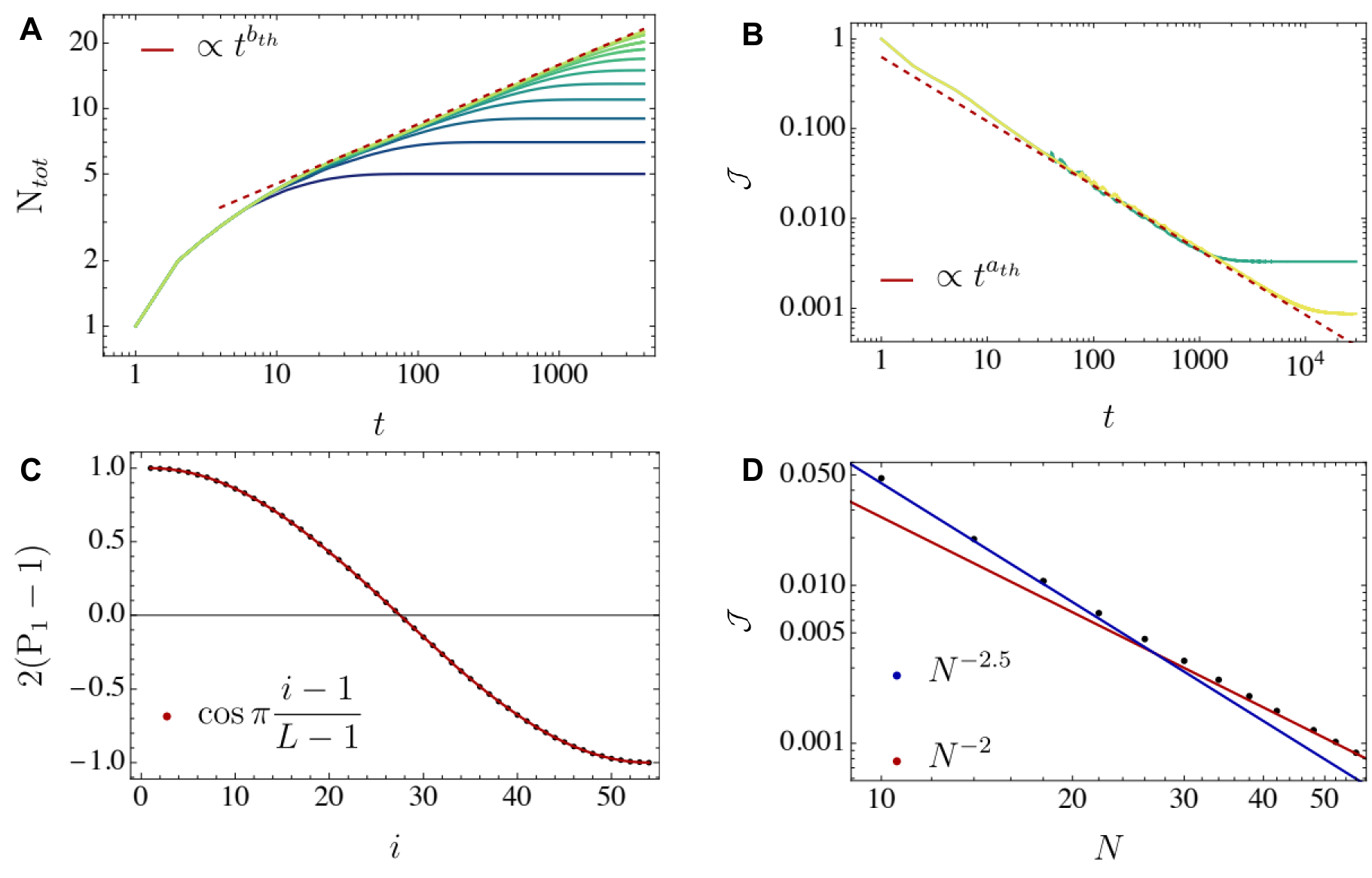}
\caption{Numerical simulations of quantum transport in the boundary-driven Floquet XXZ chain at the isotropic point $\phi = 2 \theta = \pi/2$, in the absence of external decoherence. The bond dimension of the MPDO is truncated to $\chi = 128$. (A) Time dependence of the total number of particles into the system $N_{tot} = \sum^{N-1}_{i=2}n_i$ for system sizes of $N = 10-56$ qubits (blue to yellow colors), as a function of the number of driving cycles.  We find a  power-law scaling law with an exponent $b_{th} \sim 0.2746$ which develops after an initial transient. (B) Pumping current as a function of time for $N = 30, 56$ qubits exhibits an exponent $a_{th} = -0.7178 \approx b_{th} - 1$.   (C) A normalized local particle number in the NESS as a function of qubit position for $N = 56$. The cosine function is the  strong driving limit prediction for the case of solvable boundaries, at the isotropic point of the XXZ Hamiltonian~\cite{ProPRL11}. (D) NESS current scaling with the system size. We observe that for larger system sizes $\mathcal{J} \propto N^{-2}$ while for smaller sizes the exponent is slightly larger. }
\label{fig:clean}
\end{figure}

The driving of the system is realized by a trace-preserving operation, where the auxiliary qubits are reset to a $\ket{0}$ or $\ket{1}$ state. The local quantum channel that corresponds to this operation can be formally expressed with a set of two Kraus operators,
\begin{equation}
 K^1_{1,i} = \frac{ n_i +\sigma^+_i}{\sqrt{2}} \hspace{5 mm}  K^1_{2,i} = \frac{n_i -\sigma^+_i}{\sqrt{2}} \hspace{10 mm} K^0_{1,i} = \frac{ 1 - n_i +\sigma^-_i}{\sqrt{2}} \hspace{5 mm}  K^0_{2,i} = \frac{1- n_i -\sigma^-_i}{\sqrt{2}},
\end{equation}
where,
\begin{equation}
  \sum^2_{l=1}  K^{1}_{l,i} \rho  (K^{1}_{l,i})^\dag = |1_i\rangle \langle 1_i| \text{tr}_i \rho \equiv \mathcal{K}^1_i (\rho),\hspace{10mm}  \sum^2_{l=1}  K^{0}_{l,i} \rho  (K^{0}_{l,i})^\dag = |0_i\rangle \langle 0_i| \text{tr}_i \rho \equiv \mathcal{K}^0_i (\rho),
\end{equation}
and satisfy $\sum^2_{l=1}(K^m_{l,i})^\dag K^m_{l,i} = 1_{2\times 2}$. The index $i$ denotes the auxiliary qubit $i = 1,2$ which is reset by the operation. We use the calligraphic letters to denote the action of the quantum channel on a state. We additionally denote the reset channel at both boundaries by $\mathcal{K}^{m_1 m_2} = \mathcal{K}^{m_1}_{1}\otimes \mathcal{K}^{m_2}_{2} $. Following the reset operation, we couple the auxiliary qubit to the system using swap gates,
\begin{equation}\label{eq:a-s}
    U_{B} = \text{iSWAP}_{(a,1),1}\text{iSWAP}_{(a,2),L},\hspace{10mm} \text{iSWAP}_{(a,i),j}= \text{FSim}_{(a,i),j}\left(\frac{\pi}{2} ,0\right).
\end{equation}

A stroboscopic time step, in the absence of decoherence, starts with the reset of the auxiliary qubits to states $m_1,m_2$, followed by the auxiliary-system coupling, Eq.~(\ref{eq:a-s}), and is completed by the unitary evolution according to Eq.~(\ref{eq:XXZ}),
\begin{equation}\label{eq:Reset}
    \rho(d+1) =   U_{XXZ} U_B\mathcal{K}^{m_1 m_2}\left(\rho(d)\right) U^\dag_B U^\dag_{XXZ}.
\end{equation} 

We consider an initial product state with qubits initialized in the state $\rho(0) = |0 \ldots 0_N\rangle \langle 0 \ldots 0_N|$. For the pumping protocol we will assume that the first qubit, resets to state $|1\rangle$, $m_1 = 1$, while the last qubit resets to state $|0\rangle$, $m_2 = 0$. It is evident by construction that our protocol generates maximal pumping of particles in the system, as at the start of each driving cycle the left-most auxiliary qubit, $Q_{a,1}$ is always in state $|1\rangle$. 

To study the transport properties of the system, we measure local occupations $\langle n_i\rangle$ at half-integer times (that is, after an integer number of cycles and also in the middle of cycles, see main text). This allows us to obtain the local currents. 

\begin{figure}[t]
\includegraphics[width=1\columnwidth]{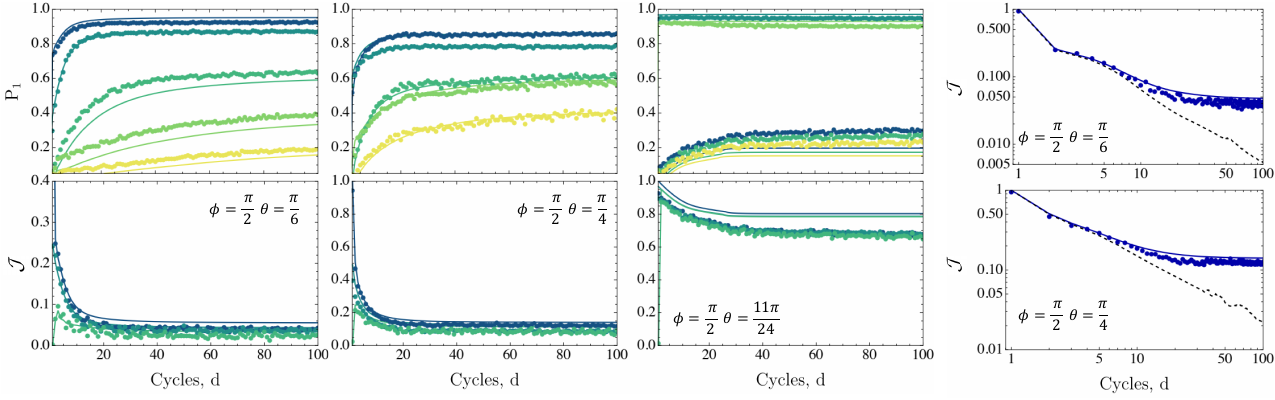}
\caption{Left three columns: A comparison between the experimental data and tensor-network simulations in the presence of weak decoherence. For the numerical simulations we used MPDO parametrization of the density matrix with bond dimension $\chi = 500$. The top 3 plots show the particle numbers of the five system qubits closest to the left auxiliary, where color varies according to position, $1 \rightarrow 5$ corresponding to blue $\rightarrow$ yellow. The bottom plots illustrate the three local currents closest to the auxiliary site. The values of decoherence are $(\gamma_\theta, \gamma_\phi) = (0.01 , 0.03),(0.016 , 0.038),(0.016 , 0.038) $ for the three values of parameters $\theta = \pi/6 , \pi/4, 11 \pi/24$, respectively. Right column: The decay of the pumping current $\mathcal{J} = \mathcal{J}_{in}$ as a function of time for different parameters. The points denote experimental data and the solid lines are the result of the simulation in the presence of the weak decoherence specified above. The dashed line shows the decoherence-free simulation. 
 }
\label{fig:decoherence}
\end{figure}

For the numerical evolution of the state, we employ tensor-network description of the density matrix known as matrix product density operator (MPDO)~\cite{VerstraetePRL04}. The time integration is performed  by a time-evolving block 
decimation (TEBD) algorithm~\cite{VidalPRL03}, implemented using ITensor library~\cite{itensor}. 

Furthermore, we model the (uncontrolled) decoherence as a product of local quantum channels,
\begin{equation}\label{eq:decoherence}
\mathcal{D} = \bigotimes^{N}_{i=1}\mathcal{D}_i,\qquad \mathcal{D}_i\left(\begin{array}{cc}
\rho_{1,1} & \rho_{1,0}\\
  \rho_{0,1}&  \rho_{0,0}
\end{array}\right)= \left(\begin{array}{cc}
e^{-\gamma_{\theta}}\rho_{1,1} & e^{-\gamma_{\phi}-\gamma_{\theta}/2}\rho_{1,0}\\
 e^{-\gamma_{\phi}-\gamma_{\theta}/2} \rho_{0,1}& (1- e^{-\gamma_{\theta}})\rho_{1,1} + \rho_{0,0}
\end{array}\right),
\end{equation}
where $\gamma_{\theta}$/$\gamma_{\phi}$ are the decay and dephasing noise rates, respectively. The local quantum channel $\mathcal{D}_i$ can be equivalently defined by jump operators $l_1 = \sqrt{\gamma_{\theta}} \sigma^-$, $l_2 = \sqrt{\frac{\gamma_{\phi}}{2} }\sigma^z$, time-integrated using the standard Lindblandian formalism over one unit of time. The decoherence map is applied to the system at the end of every Floquet step, Eq.~(\ref{eq:Reset}).

\subsection{Dynamics and steady state in the absence of decoherence at the isotropic point}

We start by analyzing the transport properties in the absence of external decoherence, focusing on the isotropic point $\phi=2\theta$. In Fig.~5 of the main text we showed the power-law temporal dependence of current through the system, with a dynamical exponent that corresponds to a subdiffusive phase. We perform large-scale tensor-network simulations to verify the presence of a sub-diffusive dynamical exponent for various system sizes and at all timescales. The results are shown in Fig.~\ref{fig:clean}A and Fig.~\ref{fig:clean}B. Here we find that the number of particles in the system follows a universal dynamical exponent $b_{th} \approx 0.2746$ and the rate of pumping particles displays a dynamical exponent $a_{th} = -0.7178 \approx b_{th} - 1$. As we will later see, the small deviation from the experimental value $a_{ex} \sim -0.64$ can be attributed to the presence of weak decoherence.  We note that the total number of particles is not directly equivalent to the rate of pumping particles. It is the time integrated difference between the rate of pumping and the rate of dissipating particles from the other end of the chain, $N_{tot} = \int_t dt\left(\mathcal{J}_{in}(t)-\mathcal{J}_{out}(t)\right)$. We have explicitly checked that for times sufficiently smaller than the saturation timescale, defined by the approach to the NESS, the outgoing current is weak ($\mathcal{J}_{in} \gg \mathcal{J}_{out}$).

Furthermore, in Fig.~\ref{fig:clean}C and Fig.~\ref{fig:clean}D, we extract the properties of the non-equilibrium steady state (NESS). We observe that for large system sizes the local current scales with the system size as $\mathcal{J}\propto N^{-2}$ and a particle profile follows a relation $\langle n_i \rangle \sim \frac{1}{2}\cos{\pi\frac{i-1}{L-1}}+1$. These predictions are in agreement with the strong driving limit prediction, for the case of solvable boundaries, at the isotropic point of the XXZ Hamiltonian~\cite{ProPRL11}. However, the dynamical exponent $a_{th}$ of the transient regime does not have a clear connection to the current exponent $c = 2$ of the NESS. The reason for that is the large deviation from linear response. However, it is worth noting that linear response arguments~\cite{LiPRL03} would lead to a relation between the two exponents $b_{LR} = \frac{1}{1+c} = 1/3$ and therefore $a_{LR} = -2/3$. Interestingly the theoretical exponents we observe are relatively close to these values.

\subsection{Effects of decoherence on the dynamics and the steady state}

In this subsection, we investigate the effects of external decoherence on transport. We use a simplified model of decoherence described by Eq.~\ref{eq:decoherence}, and assume uniform strength of decoherence across the device. In Fig.~\ref{fig:decoherence} we illustrate that across all dynamical regimes, both polarization and currents are in a good qualitative agreement with numerical simulations where dephasing noise and decay rates are chosen to be $\gamma_\phi \sim 0.03$ to $0.04$, $\gamma_\theta \sim 0.01$ to $0.015$. The agreement is worse for the ballistic regime $\theta = 11 \pi/24$. We attribute this to the fact that the steady state in this regime depends very strongly on the precise values of $\theta, \phi$. Small fluctuations of the order of $2-3\%$ are sufficient to explain the observed difference. The experimental errors on $\theta$ and $\phi$ are also larger for $\theta$ values closer to $\pi / 2$ (Fig.~\ref{fig:s2}). In addition, we explicitly illustrate that the deviation of the observed pumping current $\mathcal{J} = \mathcal{J}_{in}$ from the decoherence-free numerical results can be accurately explained by decoherence both at finite times and the NESS.

\end{document}